\documentclass[journal,10pt,twocolumn,twoside]{IEEEtran}

\usepackage{mathtools}
\usepackage{empheq}
\usepackage{algpseudocode, algorithm}
\usepackage{algorithmicx}
\usepackage{subfig}
\usepackage[table]{xcolor}%
\usepackage{varwidth}
\usepackage{url}
\usepackage{multirow}
\usepackage{amssymb}
\usepackage{mathtools}
\usepackage{changes}
\usepackage[short]{optidef}
\usepackage{caption}
\usepackage{amsmath}
\usepackage{textcomp}
\usepackage{gensymb}
\usepackage{cite}
\usepackage{booktabs}
\usepackage{hyperref}
\usepackage{balance}
\usepackage{booktabs}
\usepackage{array}
\usepackage{balance}
\newcolumntype{P}[1]{>{\centering\arraybackslash}p{#1}}
\usepackage{multirow}
\usepackage{makecell}

\begin{document}
	
	\title{6G Internet of Things: A Comprehensive Survey}
	
	\author{Dinh C. Nguyen,	Ming Ding, Pubudu N. Pathirana, Aruna Seneviratne, Jun Li, \\ Dusit Niyato,~\IEEEmembership{Fellow,~IEEE}, Octavia Dobre,~\IEEEmembership{Fellow,~IEEE}, and H. Vincent Poor,~\IEEEmembership{Fellow,~IEEE}
		
		\thanks{Dinh C. Nguyen and Pubudu N. Pathirana are with the School of Engineering, Deakin University, Waurn Ponds, VIC 3216, Australia (e-mails: \{cdnguyen, pubudu.pathirana\}@deakin.edu.au).}
		\thanks{Ming Ding is with Data61, CSIRO, Australia (email: ming.ding@data61.csiro.au).}
		\thanks{Aruna Seneviratne is with the School of Electrical Engineering and Telecommunications, University of New South Wales (UNSW), NSW, Australia (email: a.seneviratne@unsw.edu.au).}
		\thanks{Jun Li is with the School of Electrical and Optical Engineering, Nanjing University of Science and Technology, Nanjing 210094, China (e-mail: jun.li @njust.edu.cn).}
		\thanks{Dusit Niyato is with the School of Computer Science and Engineering, Nanyang Technological University, Singapore (email: dniyato@ntu.edu.sg).}
		\thanks{Octavia Dobre is with the Faculty of Engineering and Applied Science, Memorial University, Canada (e-mail: odobre@mun.ca).}
		\thanks{H. Vincent Poor is with the Department of Electrical and Computer Engineering, Princeton University, Princeton, NJ 08544 USA (e-mail: poor@princeton.edu).}
		
		\thanks{This work was supported in part by the CSIRO Data61, Australia, and in part by U.S. National Science Foundation under Grant CCF-1908308. The work of Jun Li was supported by National Natural Science Foundation of China under Grant 61872184. }
	
	}

	\markboth{}%
	{}

	\maketitle
	
	\begin{abstract}
	The sixth generation (6G) wireless communication networks are envisioned to revolutionize customer services and applications via the Internet of Things (IoT) towards a future of fully intelligent and autonomous systems. In this article, we explore the emerging opportunities brought by 6G technologies in IoT networks and applications, \textcolor{black}{by conducting a holistic survey} on the convergence of 6G and IoT.  \textcolor{black}{We first shed light on some of the most fundamental 6G technologies that are expected to empower future IoT networks,} including edge intelligence, reconfigurable intelligent surfaces, space-air-ground-underwater communications, Terahertz communications, massive ultra-reliable and low-latency communications, and blockchain. Particularly, compared to the other related survey papers, we provide an in-depth discussion of the roles of 6G in a wide range of prospective IoT applications via five key domains, namely Healthcare Internet of Things, Vehicular Internet of Things and Autonomous Driving, Unmanned Aerial Vehicles, Satellite Internet of Things, and Industrial Internet of Things. Finally, we highlight interesting research challenges and point out potential directions to spur further research in this promising area.
	
	\end{abstract}
	
	
	\IEEEpeerreviewmaketitle

\section{Introduction}

Recent advances in wireless communications and smart device technologies have promoted the proliferation of Internet of Things (IoT) with ubiquitous sensing and computing capabilities to interconnect millions of physical objects to the Internet. Nowadays, IoT constitutes an integral part of the future Internet and has received much attention from both academia and industry due to its great potential to deliver customer services in many aspects of modern life \cite{1}. IoT enables seamless communications and automatic management between heterogeneous devices without human intervention which has the potential to revolutionize industries and provide significant benefits to society through fully intelligent and automated remote management systems. 

\begin{figure*}
	\centering
	\includegraphics[width=0.94\linewidth]{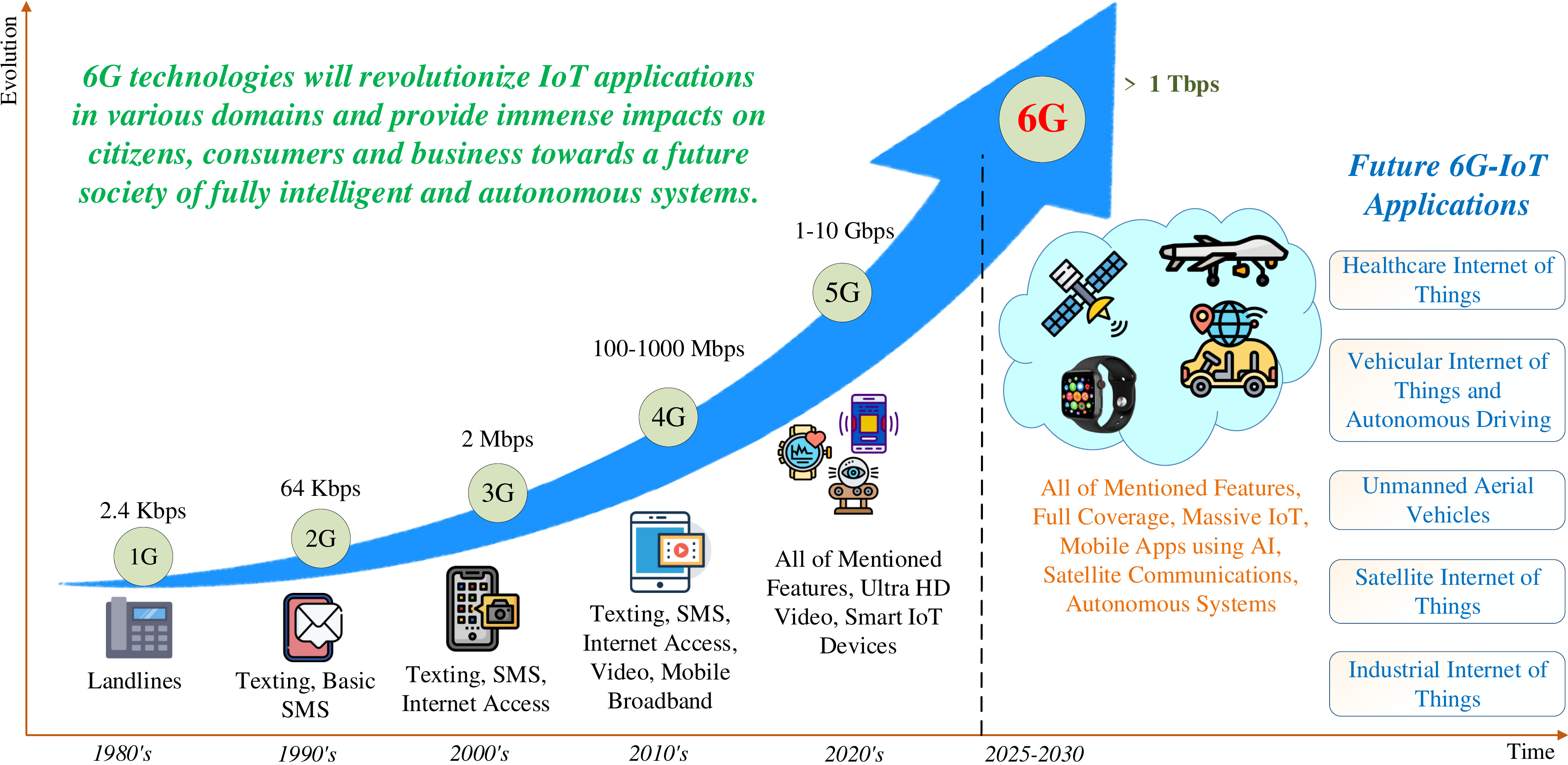}
	\caption{The evolution of wireless networks toward future 6G-IoT. }
	\label{Fig:6G-IoT}
	\vspace{-0.1in}
\end{figure*}

As an enabler for supporting IoT networks and applications, mobile technologies from the first to the fifth generation have been already proposed and deployed commercially, as illustrated in Fig.~\ref{Fig:6G-IoT}. Notably, enabled by inherent usage features such as enhanced mobile broadband (eMBB), massive machine type communication (mMTC), and ultra-reliable and low-latency communication (URLLC) services, the latest fifth-generation (5G) technology has been proven to offer different service opportunities to IoT ecosystems with high throughput, low latency and energy-efficient service provision \cite{2}, \cite{205}. However, with the unprecedented proliferation of smart devices and the rapid expansion of IoT networks, 5G cannot completely meet the rising technical criteria, e.g., autonomous, ultra-large-scale, highly dynamic and fully intelligent services. The fast growth of automated and intelligent IoT networks is likely to exceed the capability of the 5G wireless systems. Moreover, the emergence of new IoT services and applications such as remote robotic surgery and flying vehicles, also requires further advances in current 5G systems for improving the quality of IoT service delivery and business \cite{3}. 

\begin{figure*}
	\centering
	\includegraphics[width=0.97\linewidth]{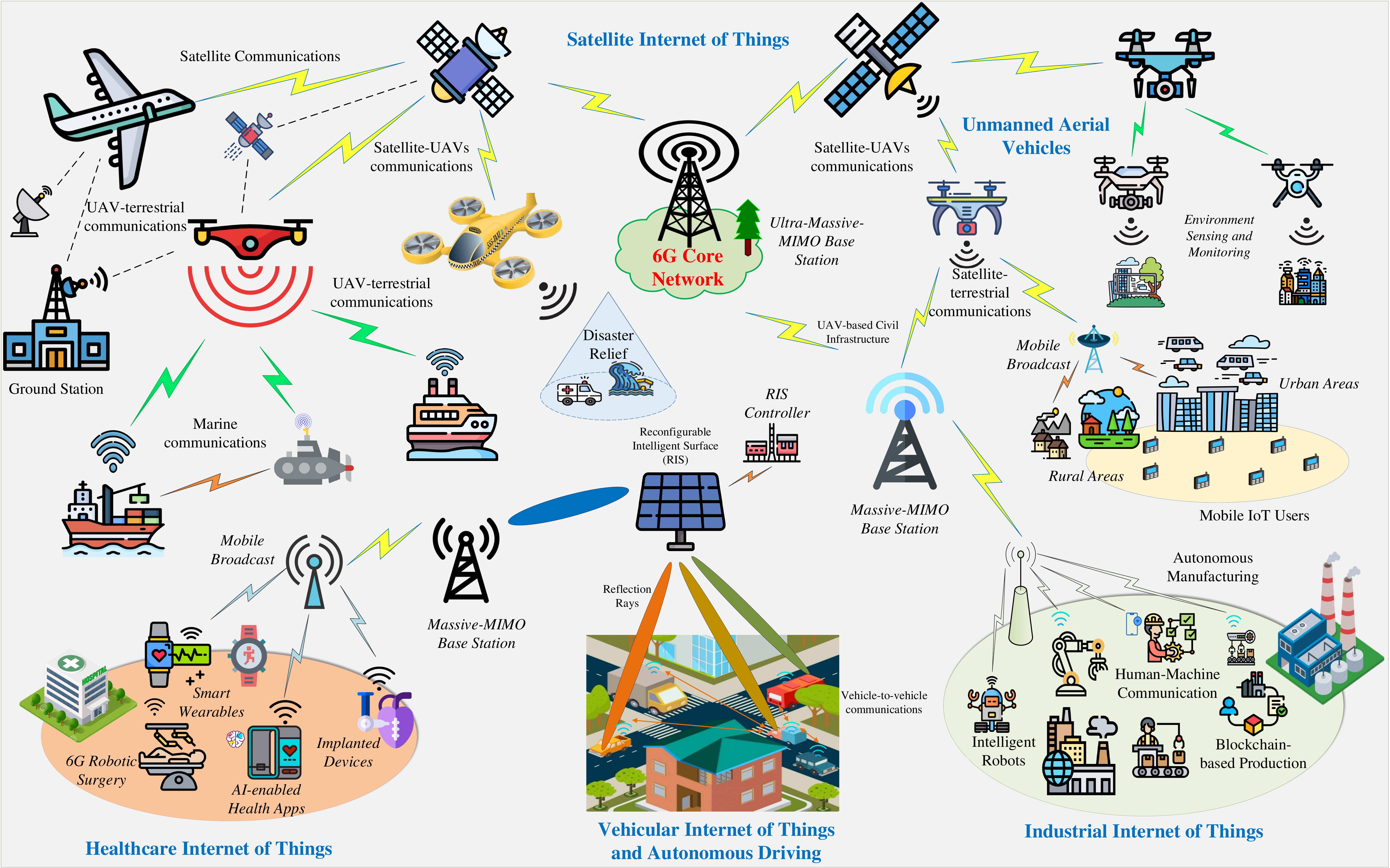}
	\caption{Vision of future 6G-based IoT applications. }
	\label{Fig:Overall}
	\vspace{-0.1in}
\end{figure*}

To pave the way for the development in IoT and beyond, research on sixth-generation (6G) wireless networks \cite{4} and their accompanying technological trends has recently received much attention from both academia and industry. 6G is expected to provide an entirely new service quality and enhance user's experience in current IoT systems due to its superior features over the previous network generations, such as ultra low-latency communications, extremely high throughput, satellite-based customer services, massive and autonomous networks \cite{7,8,add1poor}. These levels of capacity will be unprecedented and will accelerate the applications and deployments of 6G-based IoT networks across the realms of IoT data sensing, device connectivity, wireless communication, and 6G network management. Enabled by the great potential of 6G-IoT, many efforts have been put into research in this promising area. For instance, Finland has sponsored the first 6G project named 6Genesis \cite{4}, and built the world's first experimental 6G-IoT research environment. Nokia has launched Hexa-X \cite{5}, a new European 6G flagship research initiative from January 1, 2021, aiming to develop the vision for future 6G systems for connecting human, physical and digital worlds in future IoT networks through the collaboration of prominent European network vendors, communication service providers, and research institutes. Furthermore, the U.S. Federal Communications Commission (FCC) has opened the Terahertz (THz) spectrum band which allows researchers and engineers to test 6G functions on mobile communications systems and IoT devices \cite{6}. 
\textcolor{black}{Moreover, the government of South Korea has planned to launch a pilot project for 6G mobile services from 2026 \cite{Korea}. In this project, five major IoT areas have been selected for testing and evaluation of 6G systems, including digital healthcare immersive content, self-driving cars, smart cities and smart factories. Initial 6G networks could be deployed in 2028, while mass commercialization of this technology is expected to occur in 2030.} These recent activities have motivated researchers to look into the significant promise of 6G-IoT and exploit fundamental technologies for enabling future 6G-IoT, aiming to satisfy the requirements for the intelligent information society of 2030s. The vision of 6G-IoT applications that will be discussed in this paper is illustrated  in Fig.~\ref{Fig:Overall}. 
\subsection{Comparison and Our Key Contributions}
Driven by the recent advances of wireless networks and IoT, some research efforts have been made to review related works. Specifically, the study in \cite{9} provided a brief discussion on the potential technologies for 6G to enable mobile AI applications and the analysis of the AI-enabled solutions for 6G network design and optimization. A speculative study on 6G was given in \cite{10} where the authors indicated the visionary technologies potentially used in future 6G networks and applications. The use cases in 6G wireless networks were summarized in \cite{11}, while the recent advances in wireless communication toward 6G were presented in \cite{12}. The works in \cite{13}, \cite{14} presented a survey on the vision of future 6G wireless communication and its network architecture, with a focus on the analysis of enabling technologies for 6G networks. A more comprehensive survey on 6G wireless communications and networks was provided in \cite{15}, where the authors paid attention to the illustration of the tentative roadmap of definition, specification, standardization, and regulation in 6G technologies. The potential 6G requirements and the latest research activities related to 6G were also discussed in \cite{16}, while the survey in \cite{17} provided a holistic discussion of various essential technologies in 6G. The comparison of the related
works and our paper is summarized in Table~\ref{Table:Comparisons}. 

\textcolor{black}{Although 6G has been studied extensively in the literature, there is no existing work to provide a comprehensive and dedicated survey on the integration of 6G and IoT, to the best of our knowledge. Notably, a holistic discussion on the emerging 6G-IoT applications such as Vehicular Internet of Things, Autonomous Driving, and Satellite Internet of Things is still missing in the open literature.} These limitations motivate us to conduct a holistic review on the convergence of 6G and IoT. Particularly, we identify and discuss the most fundamental 6G technologies for enabling IoT networks, including edge intelligence, reconfigurable intelligent surfaces, space-air-ground-underwater communications, THz communications, massive URLLC communications, and blockchain. \textcolor{black}{It is worth noting that while the existing works \cite{11,12,13,14,15} only focus on the discussion of fundamental technologies for enabling wireless communications and networks, we here highlight the 6G technologies which directly influence the IoT applications and network in a holistic manner, from intelligence (e.g., edge intelligence), communications (e.g., massive URLLC communications) to security (e.g., blockchain).} The representative use cases on the integration of 6G technologies and IoT are also explored and analyzed. \textcolor{black}{Subsequently, we present an extensive discussion on the use of these 6G technologies in a wide range of newly emerging IoT applications via five domains, i.e., Internet of Healthcare Things, Unmanned Aerial Vehicles, Vehicular Internet of Things and Autonomous Driving, Satellite Internet of Things, and Industrial Internet of Things. The taxonomy tables are also provided to give more insights into the convergence of 6G and IoT.} Finally, we discuss a number of important research challenges and highlight interesting future directions in 6G-IoT. \textcolor{black}{These technical contributions thus make our article fundamentally different from the existing survey papers in the open literature.} In a nutshell, this article brings a new set of contributions as highlighted below: 
\begin{enumerate}
	\item \textcolor{black}{We present a holistic discussion of the convergence of 6G and IoT,} starting from an introduction to the recent advances in 6G and IoT and the discussion of the technical requirements of their integration.
	\item	\textcolor{black}{We extensively discuss the fundamental 6G technologies which are envisioned to enable IoT networks,} including edge intelligence, reconfigurable intelligent surfaces, space-air-ground-underwater communications, THz communications, massive URLLC, and blockchain.
	\item	We then provide an extensive survey and discussion of \textcolor{black}{the roles of 6G in prospective IoT applications in five key domains}, namely Healthcare Internet of Things, Unmanned Aerial Vehicles, Vehicular Internet of Things and Autonomous Driving, Satellite Internet of Things, and Industrial Internet of Things. \textcolor{black}{The representative use cases in each 6G-IoT application domain are highlighted and discussed.} Moreover, taxonomy tables to summarize the key technical aspects and  contributions of each 6G-IoT use case are also provided.
	\item	We identify several important research challenges and then discuss possible directions for future research toward the full realization of 6G-IoT.
\end{enumerate}

\subsection{Structure of The Survey}
\begin{figure*}
	\centering
	\includegraphics[width=0.99\linewidth]{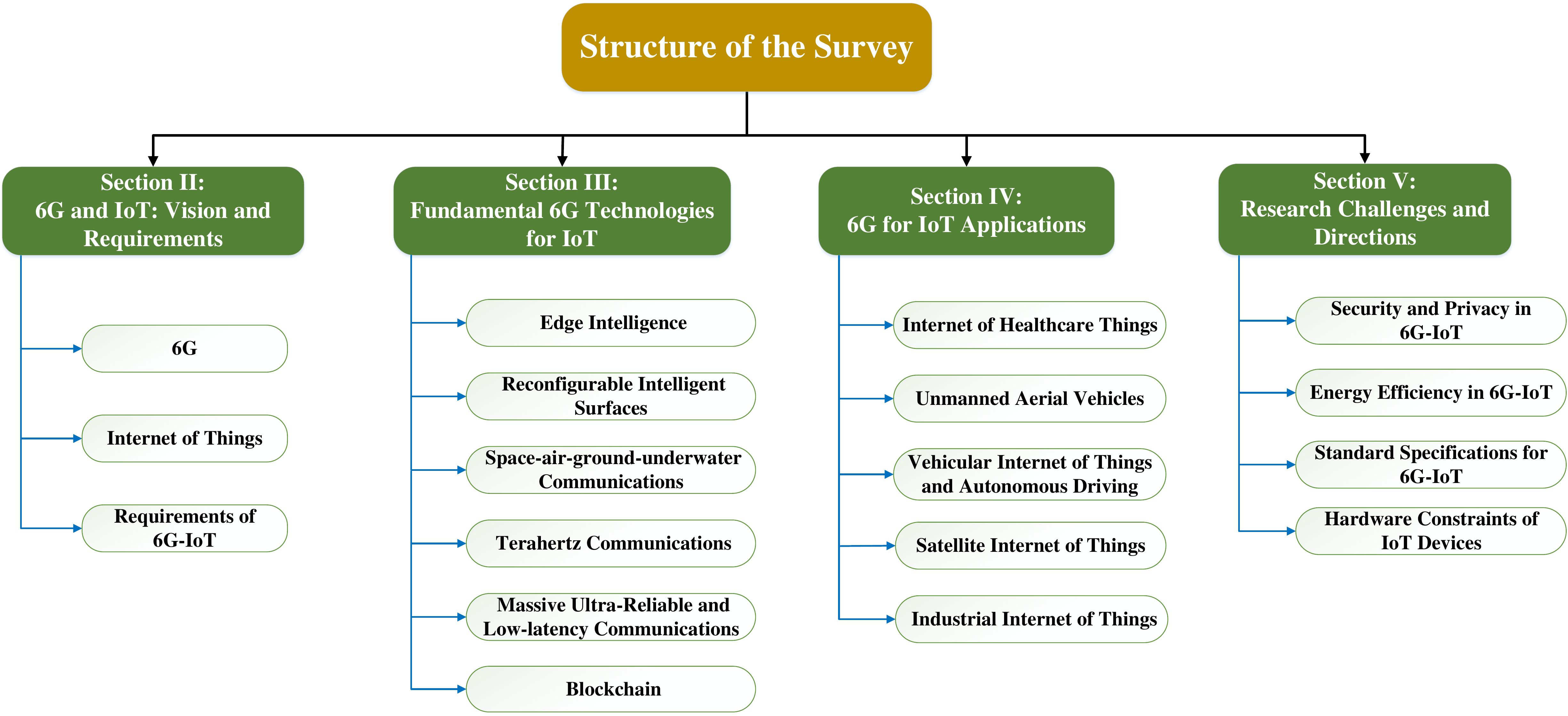}
	\caption{\textcolor{black}{Organization of this article.} }
	\label{Fig:0_Structure}
	\vspace{-0.1in}
\end{figure*}

This survey is organized as shown in Fig.~\ref{Fig:0_Structure}.  Section~\ref{Sec:State-of-Art} discusses the recent advances and vision of 6G and IoT and highlights the requirements of their integration. Next, the fundamental technologies enabling the 6G-IoT networks and applications are analyzed in Section~\ref{tech}. The opportunities brought by 6G in a number of newly emerging IoT applications are explored and discussed in Section~\ref{app} in a number of important domains, i.e., Healthcare Internet of Things (HIoT), Unmanned Aerial Vehicles (UAVs), Vehicular Internet of Things (VIoT) and Autonomous Driving, Satellite Internet of Things (SIoT), and Industrial Internet of Things (IIoT). Section~\ref{cha} identifies several key research challenges, including security and privacy in 6G-IoT, energy efficiency in 6G-IoT, hardware constraints of IoT devices, and standard specifications for 6G-IoT, along with the discussion of possible directions for future research. Finally, Section~\ref{concl} concludes the article. A list of key acronyms used throughout the paper is summarized in Table~\ref{Table:Acronyms}.

\begin{table*}
	\centering
	\caption{\textcolor{black}{Existing surveys on 6G-related topics and our new contributions.  }}
	\label{Table:Comparisons}
		\setlength{\tabcolsep}{5pt}
		\begin{tabular}{|P{1.3cm}|P{4cm}|P{11.5cm}|}
			\hline
			\centering \textbf{Related works}& 
			\centering \textbf{Topic}&	
			\textbf{Key contributions}
			\\
			\hline
			\cite{9} &	AI-based 6G concept &	A brief discussion on the potential technologies for 6G to enable mobile AI applications and the analysis of the AI-enabled solutions for 6G network design and optimization.
			\\
			\hline
			\cite{10} &	Speculative study on 6G &	A short discussion to speculate on the visionary technologies for enabling 6G networks and applications.
			\\
			\hline
			\cite{11} &	6G technologies and case studies &	A short discussion on the enabling technologies for 6G and potential case studies. 
			\\
			\hline
			\cite{12} &	6G concept &	A survey on the concept of 6G and its recent advancements in wireless communication systems. 
			\\
			\hline
			\cite{13}, \cite{14} &	Enabling technologies for 6G wireless communications &	A survey on the vision of future 6G wireless communication and its network architecture, with focus on the analysis of enabling technologies for 6G networks.
			\\
			\hline
			\cite{15} &	Technical aspects of 6G &	A discussion on the definition, specification, standardization, and regulation of 6G technologies. 
			\\
			\hline
			\cite{16} &	Requirements and research activities in 6G &	A survey on the 6G requirements and the latest research activities related to 6G. 
			\\
			\hline
			\cite{17} &	Essential technologies in 6G &	A holistic discussion of various essential technologies in 6G communications and networks.			
			\\
			\hline
			\textit{Our paper} &	6G and IoT &	A comprehensive survey on the convergence of 6G and IoT. Particularly, 
			\textcolor{black}{\begin{itemize}
				\item 	\textcolor{black}{We identify and discuss the fundamental technologies that are envisaged to enable 6G-IoT networks}, namely edge intelligence, RISs, space-air-ground-underwater communications, THz communications, massive URLLC, and blockchain.
				\item 	\textcolor{black}{We provide a holistic discussion on the emerging applications of 6G in IoT,} i.e., Healthcare Internet of Things, Unmanned Aerial Vehicles, Vehicular Internet of Things and Autonomous Driving, Satellite Internet of Things, and Industrial Internet of Things. 
				\item 	Taxonomy tables are provided to give more insights into 6G-IoT use cases. Research challenges and directions are also highlighted. 
			\end{itemize}}
			\\
			\hline
	\end{tabular}
\end{table*}

\begin{table}
	\caption{List of key acronyms.}
	\label{Table:Acronyms}
	\scriptsize
	\centering
	\captionsetup{font=scriptsize}
	\setlength{\tabcolsep}{5pt}
	\begin{tabular}{P{1.5cm}|p{5cm}}
		\hline
		\textbf{Acronyms}& 
		\textbf{Definitions}
		\\
		\hline
		6G& Sixth-Generation
		\\
		IoT	& Internet of Things
		\\
		VIoT	&	Vehicular Internet of
		Things
		\\
		HIoT& Healthcare Internet of Things
		\\
		SIoT	&	Satellite Internet of
		Things
		\\
		IIoT	&	Industrial Internet of Things
		\\
		AI& Artificial Intelligence
		\\
		FL& Federated Learning
		\\
		ML & Machine Learning
		\\
		DL & Deep Learning
		\\
		DRL & Deep Reinforcement Learning 
		\\
		DNN &Deep Neural Network
		\\
		CNN & Convolutional Neural Network
		\\
		MEC & Mobile Edge Computing 
		\\
		UAV &  Unmanned Aerial Vehicle
		\\
		URLLC	&	Ultra-Reliable Low-Latency Communication
		\\
		mURLLC	&	Massive URLLC
		\\
		THz	&	Terahertz
		\\
		RIS	&	Reconfigurable Intelligent Surface
		\\
		MIMO 	&	Multiple-Input and Multiple-Output
		\\
		NOMA	&	Non-orthogonal Multiple Access
		\\
		LEO	&	Low Earth Orbit
		\\
		V2V & Vehicle-to-vehicle
		\\
		V2X	&	Vehicle-to-Everything 
		\\
		AV	&	Autonomous Driving
		\\
		RSU	&	Road Side Unit
		\\
		QoS 	&	Quality-of-Service
		\\
		\hline
	\end{tabular}
	\label{tab1}
\end{table}

\section{\textcolor{black}{6G and Internet of Things: Vision and Requirements}}
\label{Sec:State-of-Art} 
In this section, we discuss the vision of 6G and IoT. The requirements of 6G-IoT networks are also highlighted. 
\subsection{6G}
Driven by the unprecedented proliferation of mobile devices and the exponential growth of mobile traffic, wireless communication technologies have rapidly developed in recent years as a key enabler for future customer services and applications. \textcolor{black}{Although the 5G network has been proven to enhance QoS over previous generations, it will be challenging to fully meet the newly emerging requirements of future IoT services \cite{13}. More specifically, it can be foreseen that  5G networks will be unable to  accommodate the tremendous volume of mobile traffic in 2030 and beyond. Due to the popularity of rich-video applications, enhanced screen resolution, machine-to-machine (M2M) communications, mobile edge services, etc., the global mobile traffic will grow exponentially, up to 5016 exabyte (EB) per month in the year of 2030 compared with 62 EB per month in 2020 \cite{15}. The traffic demand per mobile broadband (MBB) also increases rapidly due to the growth of smartphones and tablets and the proliferation of mobile data services. For example, there is a need for new mobile communication technologies to support video services such as Youtube, Netflix, and recently Tik-Tok since  video data traffic accounts for two thirds of all mobile traffic nowadays and continues to grow in the coming years \cite{15}. Moreover, the rapid development of data-centric intelligent systems exposes new latency limitations of 5G wireless systems. For example, the 5G air interface delay of less than 1 millisecond is inadequate to support haptic Internet-based applications such as autonomous driving or real-time healthcare assistance because the required delay is below 0.1 millisecond. }  

In this context, 6G is envisioned to provide new disruptive wireless technologies and innovative networking infrastructures to realize a plethora of new IoT applications by satisfying such stringent network demands in a holistic fashion, compared to its 5G counterpart \cite{16}. With the advent of advanced technologies such as edge intelligence, THz, and large-scale satellite constellation, 6G communication systems are able to evolve towards a more powerful IoT ecosystem as well as build the fully connected and intelligent digital world toward the foreseen economic, social, and environmental ecosystems of the 2030 era. 6G is expected to outperform 5G in multiple specifications as shown in Table~\ref{Table:6G-IoT-5G}. While 5G networks remain some critical limitations in terms of mobile traffic capability, density of device connectivity, and network latency, 6G is able to bring a new level of network qualities with the outstanding features as follows:
\begin{itemize}
	\item 	Achieving a supper high data rate from 1 Tb/s to address the massive-scale IoT connectivity where the seamless mobility, spectrum availability, and mobile traffic co-exist.
	\item 	Increasing the mobile traffic capability up to 1 $Gb/s/m^2$ to satisfy super high throughput requirements and IoT device density.
	\item 	Achieving an extremely high device connectivity density of $10^7$ devices/$km^2$ which supports well for massively-dense IoT network deployments
	\item 	Achieving ultra-low network latencies (10-100 $\micro$s) to fulfill the requirements of haptic applications, such as e-health and autonomous driving. 
\end{itemize}
\begin{table}
	\centering
	\caption{\textcolor{black}{New features of 6G-IoT versus 5G-IoT.   }}
	\label{Table:6G-IoT-5G}
	\setlength{\tabcolsep}{5pt}
	\begin{tabular}{|P{1.4cm}|P{3.1cm}|P{3.1cm}|}
		\hline
		\centering \textbf{}& 
		\centering \textbf{5G-IoT}&	
		\textbf{6G-IoT}
		\\
		\hline
		\multirow{15}{*}{\makecell{\textbf{Network} \\\textbf{Features}}}  &	\begin{itemize}
			\item	Data Rate of 20 Gb/s
			\item	Mobile Traffic Capability: 10 $Mb/s/m^2$
			\item	Connectivity Density: $10^6$ devices/$km^2$
			\item	Network Latency: 1 ms
			\item  Coverage Percentage: about 70 \%
			\item  Energy Efficiency: 1000x relative to 4G
			\item  Spectrum Efficiency: 3-5x relative to 4G
		\end{itemize} &	
		\begin{itemize}
			\item  Data Rate from 1 Tb/s 
			\item 	Mobile Traffic Capability: 1~$Gb/s/m^2$
			\item 	Connectivity Density: $10^7$ devices/$km^2$
			\item 	Network Latency: 10-100 $\micro$s
			\item  Coverage Percentage: $>$~99 \%
			\item  Energy Efficiency: 10x relative to 5G
			\item  Spectrum Efficiency: $>$ 3x relative to 5G
		\end{itemize}
		\\
		\hline
		
		\multirow{11}{*}{\makecell{\textbf{Enabling} \\ \textbf{Technologies}}} &	\begin{itemize}
			\item	mm-Wave Communications
			\item   URLLC
			\item	NOMA
			\item	Artificial Intelligence
			\item 	Cloud/Edge Computing
			\item	Software-Defined Networking/ Network Slicing
		\end{itemize} &	
		\begin{itemize}
			\item	THz Communications
			\item   Massive URLLC
			\item 	Space-air-ground-underwater Communications
			\item	Edge Intelligence
			\item	RIS
			\item	Blockchain
		\end{itemize}
		\\
		\hline
	\end{tabular}
\end{table}

\subsection{Internet of Things (IoT)}
As a key technology in integrating heterogeneous electronic devices with wireless systems, IoT aims to connect different things to the Internet, forming a connected environment where data sensing, computation and communications are performed automatically without human involvement. IoT data can be collected from ubiquitous mobile devices such as sensors, actuators, smart phones, personal computers, and radio frequency identifications (RFIDs) to serve end users \cite{204}. It is estimated that IoT will achieve an impressive development in the next few years.   According to Cisco \cite{18}, up to 500 billion IoT devices are expected to be connected to the Internet by 2030, from only 26 billion devices in 2020. Furthermore, in a new analysis of IHS Markit \cite{19}, a world leader in critical information, analytics and solutions, the development of connected global IoT devices will achieve an impressive rate of 12 percent annually, from nearly 27 billion in 2017 to 125 billion in 2030. Recently, GlobeNewswire also forecasts that the global 5G-IoT market will grow from USD 694.0 million in 2020 to USD 6,285.5 million by 2025 \cite{20}. \textcolor{black}{The Internet of Nano-Things is also significant to build future advanced IoT ecosystems \cite{400}, where the network of objects (nano-devices and things) can sense, transmit, process, and store data based on  nano units (e.g., a nanocontroller) for supporting customer services such as healthcare monitoring. Seamless interconnectivity among nano-networks via the available communication networks and the Internet requires developing new network architectures and new communication paradigms.} In this context, 6G with its exceptional features and strong capabilities will be a key enabler for supporting future IoT networks and applications, by providing full-dimensional wireless coverage, and integrating all functionalities, from sensing, communication, computing, to intelligence and fully autonomous control. In fact, the next generation 6G mobile networks are envisaged to provide massive coverage and better scalability to facilitate IoT connectivity and service delivery, compared to the 5G mobile network \cite{21}. 
\subsection{Requirements of 6G-IoT}
To fully realize the 2030 intelligent information society of full intelligence, massive device connectivity and coverage, data-driven services, and autonomous systems, 6G-IoT will need more stringent requirements over its 5G-IoT counterpart, as highlighted below.
\subsubsection{Massive IoT Connectivity}
Driven by the explosion of smart devices and rapid development of wireless communication technologies, the mobile connectivity has increased tremendously \cite{13}. It is predicted that the volume of global mobile traffic will grow exponentially with over 5000 exabyte in the year of 2030 and increase 80 times compared to the mobile traffic in 2020.  Broadband access platforms along with satellite networks enabled by using low earth orbit (LEO) satellites \cite{22} will be the key enabler for supporting large-scale communications of future smart devices.  Furthermore, the use of flying platforms such as UAVs are also needed to support seamless connectivity over the future large-scale IoT networks where fixed base stations cannot manage to ensure stable and reliable device communications in moving IoT networks. \textcolor{black}{For example, a distributed UAV deployment prototype is presented in \cite{23} in the context of 6G networks, by using 33 UAVs to provide wireless connectivity for 400 terrestrial IoT users over the coverage of 2000 m x 2000 m area. Given the UE distribution and corresponding dis-continuous UAV location space, a distributed motion algorithm is developed which allows each UAV to autonomously obtain the optimal position in a continuous IoT space. Simulations indicate that the use of swarm of UAVs is able to obtain the maximum load balance near to 1 by the distributed deployment solution, with the deployment time reduced by up to 40\% compared to the centralized method.}

\subsubsection{Massive Ultra-Reliable Low-Latency IoT Communications}
 Although ultra-reliable and low-latency communication (URLLC) has been introduced and used in applied 5G-based IoT use cases \cite{24}, it needs to be improved to massively support emerging applications in 6G-IoT networks such as fully autonomous IoT and flying IoT systems. For example, in the future autonomous transportation systems, where vehicles are self-controlled and navigated in real-time, the massive URLLC is highly necessary for transporting video feeds from cameras to vehicles and coordinating the timely vehicle signalling on the roads in an automated and safe manner. In the future, timeliness of information delivery will be a significant feature for the intelligent interconnected society where the tactile internet will dominate to offer haptic communications for mission-critical IoT services with touch and actuation in real-time. 
\textcolor{black}{As an case study, the work in \cite{300} implements a massive URLLC-based IoT simulation for 6G IoT networks by designing a  short range wireless isochronous real-time in-X subnetwork with communication cycles shorter than 0.1 ms and outage probability below $10^{-6}$. A dense IoT scenario is considered with up to two devices per $m^2$ using a multi-GHz spectrum for providing high spatial service availability. By conducting a semi-analytical system evaluation analysis, it is revealed that the cycle times are a factor of x10 shorter than the latency targets of 5G radio technologies (i.e., below 0.1 ms) which is potentially applied to future ultra-low latency IoT applications such as autonomous driving, real-time health monitoring, and industrial automation. }

\subsubsection{Improved 6G IoT Communication Protocols}
The introduction of new vertical IoT applications in future intelligent networks imposes major architectural changes to current mobile networks in order to simultaneously support a variety of stringent requirements (e.g., autonomous driving and e-healthcare). In such a context, network communication standards and protocols play an important role in deploying 6G-IoT ecosystems at large-scale due to the  integration with other important computing services such as edge/cloud computing and wireless technologies. For example, the Industry Specification Group of the European Telecommunications Standards Institute  has released the initiative called ETSI Multi-access Edge Computing \cite{25}, which aims to leverage seamlessly edge computing and communication frameworks for IoT-based applications originating from vendors, developers and third-party service providers. This initiative could be a significant step to the deployment of future IoT applications at the network edge in the 6G network where IoT computing and storage are expected to be shifted from the network centre to the network edge. \textcolor{black}{Recently, the IEEE 802.11 working group has initiated discussions on releasing the next generation of Wi-Fi standard, referred to as IEEE 802.11be Extremely High Throughput \cite{26}, which can meet the peak throughput requirements set by upcoming IoT applications in the 6G era. These communication standards are expected to support well the service providers in deploying intelligent IoT services at the network edge. }
\subsubsection{Extended IoT Network Coverage}

In the future IoT-based society, it is desirable to achieve a full coverage beyond the terrestrial networks, from using large-dimensional space-air-ground-underwater networks \cite{4}. Edge intelligence and UAVs are the keys to achieving full wireless coverage where the former is able to provide autonomous and intelligent solutions at the network edge, while the latter can be used to build flying base stations to extend the coverage of current mobile networks from only 2D in existing terrestrial networks to 3D in an integrated terrestrial-satellite-aerial system. For example, high-altitude UAVs can be exploited as agile aerial platforms to enable on-demand maritime coverage \cite{27}, as a promising solution to establish  shore-based terrestrial based stations to facilitate the deployment of communications infrastructure for vessels on the ocean. 
\textcolor{black}{To be clear, a practical automatic identification system is employed to obtain the vessel distribution with the offshore distance of the range of [20, 30] km. To provide an on-demand coverage of distributed vessels, UAVs are coordinated to move along with vessels for ensuring long-term broadband services. Here, an oil-powered fixed-wing UAV is adopted which can perform a 740 km round trip over the coastal area of about 370 km. Once the data transmission task at a vessel is accomplished, the UAV flies back to the charging station and continues serving the next vessel user in the queue. This communication pattern is unique from traditional UAV networks where IoT users are often fixed or have random distributions.}

\subsubsection{Next-generation Smart IoT  Devices}
\textcolor{black}{Future 6G-IoT networks are predicted to rely on smart devices where edge intelligence and computing can be fully realized at the devices, e.g., smartphones, vehicles, machines and robots. Such a device-centric network poses new challenges and requirements on the design and operation of its wireless communications, since smart devices will not only generate or exploit data, but will also actively join the network management and operation processes. Device-centric wireless solutions can be device-to-device  communications or multi-hop cellular networks that are parts of the 3GPP roadmap \cite{28}. In the 6G era, each IoT smart device can act as an end-user terminal which can provide connectivity and services (e.g., intelligent control, caching, and network signalling) to other devices at the network edge without the need for a centralized controller \cite{207}. This can be extended to demand-driven opportunistic networking which is tailored to different user, service or network demands such as energy cost minimization or spectrum efficiency maximization for end-user devices \cite{29}. Moreover, recent years have witnessed a proliferation in wearable devices whose functionalities are gradually replacing the roles of smartphones that are central in 4G/5G network generations. These new wearable devices are diverse, ranging from smart wearables to smart body implants which can play a pivotal role in the revolution of wearable IoT networks for driving the emerging human-centric 6G services. }
\section{Fundamental Technologies for 6G-IoT}
\label{tech} 
\begin{figure}
	\centering
	\includegraphics[width=0.98\linewidth]{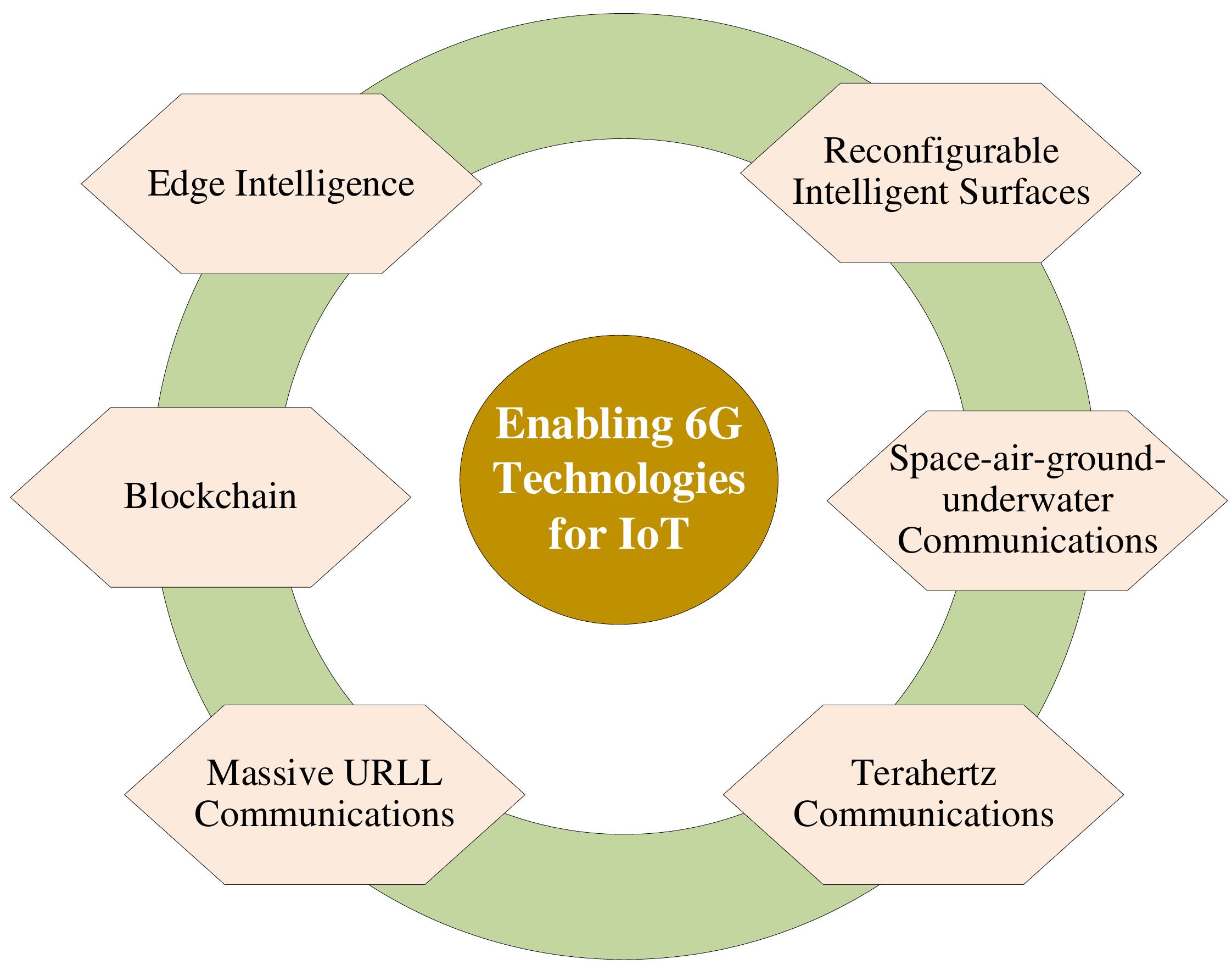}
	\caption{\textcolor{black}{Fundamental 6G technologies for IoT. }}
	\label{Fig:roadmap1}
	\vspace{-0.1in}
\end{figure}

In this section, the fundamental technologies \textcolor{black}{that are envisioned to enable future 6G-IoT networks and applications} are discussed. \textcolor{black}{It is important to highlight that existing survey papers \cite{11,12,13,14,15} mostly discuss enabling technologies for 6G wireless communication; we here focus on technologies which will directly support 6G-IoT,} including edge intelligence, RISs, space-air-ground-underwater communications, THz communications, mURLLC, and blockchain, as illustrated in Fig.~\ref{Fig:roadmap1}. 

\subsection{Edge Intelligence}
In intelligent 6G systems, AI functions are extended to the network edge thanks to the computational capabilities of edge nodes \cite{36}. This leads to a new paradigm called edge intelligence \cite{30}, \cite{31}, which is envisioned by the convergence of AI, communications, and edge computation. \textcolor{black}{Recently, some research efforts have been made to explore the potential of edge intelligence in 6G-IoT use cases. \textcolor{black}{For instance, the work in \cite{32} suggested a self-learning edge intelligence solution, aiming to identify and classify emerging unknown services from raw crowdsourcing data distributed across a wide geographical area with a case study in 6G-based vehicular networks. A latency-sensitive connected vehicular system is proposed that consists of six campus shuttles connected to two edge servers along with a mobile app to monitor the latency of wireless vehicular data transmission. A highly-efficient edge intelligence approach is employed, by equipping each vehicle with an AI-based generative adversarial network. This is able to generate synthetic data that can be directly produced by the edge server, which helps reduce the data volume required to upload from the user and mitigate the total traffic transported throughout the network. However, the proposed self-learning architecture has not yet considered user mobility that can closely affect the latency in AI training among distributed vehicles.} Moreover, the role of edge intelligence in 6G-IoT is also discussed in \cite{33} where some dominant use cases are taken into account, such as autonomous driving. Edge intelligence is foreseen as a key enabler for dynamic spectrum access to provide fast and reliable AI-based data processing at the edge devices, e.g., platoons, vehicles on the street, and road side units.} The benefits of edge intelligence are also considered for collaborative robots (or cobots) in smart manufacturing, with possible applied domains like automatic monitoring of machine health properties, autonomous or semi-autonomous navigation, and  fine-grained control of cobots. 
\textcolor{black}{Moreover, ubiquitous edge devices can run AI functions to offer large-scale edge intelligence services enabled by big data analytics \cite{402}. In this context, the information asset characterized by high volume, velocity and variety of big data is exploited to realize AI-based data analytics for its transformation into useful information to serve IoT users. Particularly, the 6G-based big data analysis technology can significantly enhance the large-scale data transmission and data computation rates based on advanced communication technologies such as massive URLLC and space-air-ground-underwater communications.}
\textcolor{black}{Along with edge computing, fog computing also allows for relieving the load on cloud servers by offering computation and storage at fog nodes close to the IoT devices for improving the QoS \cite{401}. Fog nodes would be useful to incorporate idle and spare resources of all available devices to further enhance network efficiency, especially in distributed IoT networks where cloud computing cannot handle all users' computation demands. Fog computing can be combined with AI techniques to realize fog intelligence for providing smart and low-latency IoT services in the future 6G era.} 

\begin{figure*}
	\centering
	\includegraphics[width=0.95\linewidth]{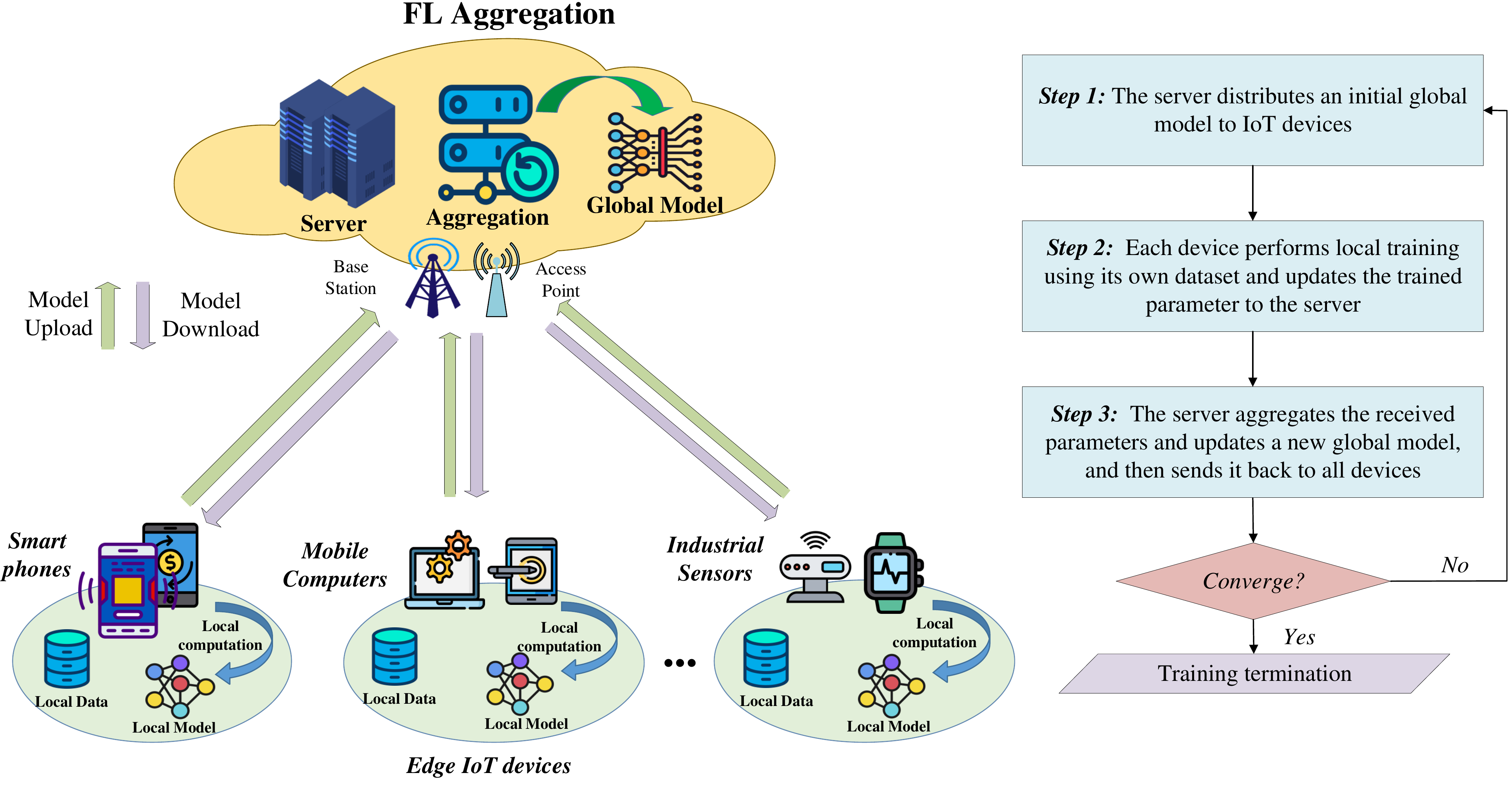}
	\caption{Edge intelligence with FL for wireless IoT networks.  }
	\label{Fig:FL-edge}
	\vspace{-0.1in}
\end{figure*}

Recently, federated learning (FL) \cite{403}  as a distributed collaborative AI approach is emerging to transform edge/fog intelligence architectures. Conceptually, FL is a distributed AI approach which enables training of high-quality AI models by averaging local updates aggregated from multiple learning edge clients without the need for direct access to the local data \cite{34}. For example, in the context of intelligent IoT networks, distributed IoT devices can collaboratively work with a data aggregator (e.g., an edge server) to perform neural network training where devices only exchange the parameters while raw data sharing is not needed, as illustrated in Fig.~\ref{Fig:FL-edge}. \textcolor{black}{With its working concept, FL can offer several unique features to IoT networks. One of the most important features is the ability to enhance data privacy based on the distributed model training without sharing raw data to external servers. Following the increasingly stringent data privacy protection legislation such as the General Data Protection Regulation (GDPR), the capability of protecting user information in FL is essential  for building future safe 6G-IoT systems. Another feature is that FL enables low-latency network communications by avoiding the offloading of huge data volumes to the remote server in the training process. This capability also helps save much network spectrum resources required for iterative data training. Additionally, the cooperation of massive IoT devices for contributing large-scale datasets and computation resources in the FL system would accelerate the convergence rate of the overall training process and thus improve learning performances, e.g., accuracy rates, for better intelligent 6G-IoT services. } \textcolor{black}{ A case study of FL-enabled edge intelligence in the 6G-IoT context is provided in \cite{35}. Motivated by the privacy leakage issues in AI training, an approach called Air-Ground Integrated Federated Learning (AGIFL) is employed to evaluate the effect of different UAV’s hovering location deployment schemes with privacy awareness. Here, multiple terrestrial nodes (e.g., mobile users) are regarded as clients to join the collaborative training with the server deployed at the UAV. In this setting, each user performs local training using its own dataset and sends the updated parameter to the UAV which aggregates all received updates to build a global model before giving it back to all participating users for the next round of training during the location deployment. By implementing simulations on image classification tasks using convolutional neural networks (CNNs) based on classical handwritten digit datasets, the proposed AGIFL-based method shows a promising classification accuracy performance (the accuracy rate achieves up to 95\%), compared to non-federated schemes in the 6G context. In the near future, the security for local gradient computation at UAVs should be taken into account in  federated training, by integrating attack detection methods such as access control and data authentication.}

\textcolor{black}{ In edge intelligence, AI on hardware will be one of the promising technologies that can accelerate the development of IoT applications in the 6G era. With the rapid growth of smart mobile devices and advancement in embedded hardware, there is an increasing interest in deploying AI on the edge devices, e.g., mobile computers, smart phones, edge servers, for both on-chip ML and DL functions. Embedded AI functions can provide on-device learning inference with low latency compared to cloud-based AI training.  The deployment of AI on hardware opens up new interesting on-device IoT applications, such as mobile smart object detection, and mobile human recognition. For example, a binary neural networks (BNNs) engine is developed in \cite{409} based on graphics processing units (GPUs) for Android devices that optimizes both software and hardware for running ML functions on resource-constrained devices, e.g., Android phones. The computation capability of BNN is exploited on mobile devices, decoupled with parallel optimizations with the OpenCL library toolkit for enabling real-time and reliable NN deployments on Android devices. The experimental results indicate that the mobile BNN architecture can achieve a learning accuracy of 95.3\%  and improve the capability of storage and CPU usage in BNN running by up to 20\%. Another case study is presented in \cite{410}, where a MobileNet app is designed for on-device CNN training on mobile devices for mobile localization applications in IoT networks. The visual images captured by phone cameras can be learned by the CNN to determine the centroid of the object (e.g., the human hand) for supporting human motion detection tasks, such as people counting in social events. }

The development of edge intelligence mostly relies on edge devices that play important roles to perform learning tasks, i.e., classification or regression. However, adversaries can deploy attacks to steal data and modify parameters during the data transmission and training, which makes edge intelligence highly vulnerable. Another security issue comes from untrusted edge devices that can exploit the private information extracted from gradient exchange in the collaborative training, which thus also introduces high risks of data breaches. Therefore, future security solutions should be proposed for a safe and reliable edge intelligence ecosystem in future 6G-IoT networks. For example, decentralizing the data learning can be a feasible choice for edge intelligence systems, where blockchain ledgers can be adopted to verify the computation process of edge intelligence models. It can be done by using allowing edge blockhain peers to authorize the accuracy of the common AI model without revealing the labels of datasets. The data flow inside the AI model can also be hidden using blockchain-based cryptography techniques to prevent data modification threats. 
\subsection{Reconfigurable Intelligent Surfaces}
Recently, reconfigurable intelligent surfaces (RISs) have gained significant research attention for 6G technology applications \cite{37}. The RIS consists of arrays of passive scattering elements with artificial planar structures, where each element is enabled by electronic circuits to reflect the impinging electromagnetic wave in a software-defined manner \cite{38}. The reconfigurability makes RISs as a promising solution to support the wireless system design and optimization by facilitating signal propagation, channel modeling and acquisition which enables smart radio environments beneficial to 6G-based applications. \textcolor{black}{Indeed, RIS enables a number of promising applications in the context of 6G-IoT networks. For instance,  RIS is able to simultaneously enhance the signal gleaned from the serving base stations in multi-cell IoT networks to reduce inter-cell interference between massive IoT devices \cite{404}. Another application of RIS is the capability to enhance the data offloading rates for IoT systems. Specifically, the volume of data offloaded to edge servers is largely dependent on the channel gain of offloading links. RISs can be deployed for establishing a virtual array gain and reflection-based beamforming gain to computation offloading links. The use of RISs thus enables more data to be offloaded to edge servers, where the data can be processed more time-efficiently than using traditional offloading approaches. }

Recent research works have started integrating RISs into 6G-based IoT applications. The study in \cite{39} provides a range of case studies where RISs can be applied to IoT such as smart buildings. Indeed, RISs can help establish the interface between the indoor and outdoor entities, aiming to facilitate the access of private households in smart buildings. From the wireless communication benefits, RISs are promising to provide cooperative entity layers for avoiding interference and improving spectral efficiencies in the device communications between indoors and outdoors. \textcolor{black}{Another case study is presented in \cite{41}, where RISs are integrated in VIoT-based vehicle-to-vehicle (V2V) networks consisting of a RIS-based access point for transmission and a RIS-based relay deployed on a building for coordinating the vehicular communications. } RISs are also considered in \cite{42} to support radio-frequency (RF) sensing for human posture recognition in IoT applications such as surveillance and remote health monitoring. By periodically programming RIS configurations in a human posture recognition system, the optimal propagation links can be obtained so that the system can create multiple independent paths that accumulate the useful information of human postures to estimate better the human posture recognitions in comparison with the random configuration and the non-configurable environment cases. 
\subsection{Space-air-ground-underwater Communications}
6G is envisioned to be a unified communication platform, not only on the land but also on the sky (such as flight) and underwater (such as voyage), to achieve an extremely broad coverage and ubiquitous connectivity for fully supporting future IoT applications. Accordingly, a cell-free and four-tier large-dimensional communication network for 6G-IoT can be derived, consisting of four tiers: space, air, terrestrial, and underwater \cite{4}.
\begin{itemize}
	\item 	\textit{Space Communication Tier:} This layer will provide wireless coverage using LEO, medium-Earth-orbit (MEO), and geostationary-Earth-orbit (GEO) satellites for areas that are not covered by terrestrial networks. Space communication technologies can be deployed to support high-capacity space networks for satellite-ground communications at various atmospheric altitudes \cite{43}.  
	\item	\textit{Air Communication Tier:} This layer is well supported by UAVs and balloons working as flying base stations to provide coverage and connectivity for managing disaster-stricken areas and supporting public safety networks and emergency situations when URLLC is required \cite{44}. UAVs also work as aerial users to incorporate with terrestrial base stations to set up direct air-ground links to perform cooperative sensing and data transmission in 6G-IoT environments. 
	\item	\textit{Terrestrial Communication Tier:} This layer aims to wireless coverage and connectivity for human activities on the ground where physical base stations, mobile devices and computing servers are interconnected together. In the context of 6G-based IoT, the THz band will be exploited to enhance spectral efficiency and accelerate communication speeds, especially in ultra-dense heterogeneous networks with millions of users \cite{45}. 
	\item	\textit{Underwater Communication Tier:} This layer accommodates connectivity services for underwater IoT devices such as submarines in broad-sea and deep-sea activities. Bi-directional communications can be necessary for interconnection between underwater IoT devices and control hubs \cite{46}. 
\end{itemize}
Several case studies have been considered for investigating the feasibility of space-air-ground-underwater communications in 6G-IoT applications. \textcolor{black}{ For example, a multiuser satellite IoT system is considered in \cite{47} based on LEO satellite communications. In this regard, a Mobile Edge Computing (MEC) server is integrated with the full-duplex access points to create satellite-links for improving communication latency efficiency in mission-critical IoT applications. The simultaneous wireless information and power transfer (SWIPT) technology is integrated with a hybrid energy storage method, i.e., the power is from power grid or renewable sources, to achieve longer battery life and higher energy efficiency for IoT communications, while the satellite provides the wide area network connection for terrestrial terminals, including stations and mobile users. Then, a satellite-terrestrial communication model is derived, aiming to maximize the achievable rate of IoT terminals. Simulations with a satellite IoT system with 10 IoT users illustrate an improvement in achievable rate of terminals by 16\% due to the joint optimization of CPU frequency, computation tasks, and terminal transmitting power. Nevertheless, the communication latency caused by packet loss checking over satellite  channels has not been considered in the proposed multiuser satellite IoT system optimization.  Another case study is explored in \cite{48}, where UAVs are employed to realize UAV-to-Everything communications for supporting different data transmission solutions in IoT data sensing, emergency search and monitoring, and video streaming in wireless 6G scenarios. }
	
\subsection{Terahertz (THz) Communications}
THz communications are envisioned as a driving technology for 6G-IoT which requires 100+ Gbps data rates and 1- millisecond latency. Enabled from the mm-wave spectrum, the THz band (0.1-10 THz) is promising to fulfill the future requirements of 6G-IoT applications, including pico-second level symbol duration, integration of thousands of submillimeter-long antennas, and weak interference without full legacy regulation \cite{50}. One of the key benefits of  THz spectrum is to deal with the spectrum scarcity problems in wireless communications and significantly enhance wireless system capacities in 6G-IoT. Moreover, THz communications also provide ultra-high bandwidth and high throughput, which support  ultra-broadband applications such as virtual reality and wireless personal area networks. Inspired by the unique features of THz communications, several IoT studies have been implemented in the context of 6G. Recently, THz communication technologies have been used for user localization by exploiting the ultra-wide bandwidths available at THz frequencies \cite{52}. This allows the receivers to address spaced multipath components and effectively measure the signal sent by the transmitters for estimating correctly the user location. THz bands are also exploited for UAVs communications \cite{52}, aiming to analyze the coverage probability of UAV networks with respect to the THz base station density and the strength of THz signals. \textcolor{black}{The work in \cite{53} focuses on modelling THz communication channels in VIoT-based vehicular networks  and configurations of hybrid beamforming subarray by taking transmit power consumption into account in different scenarios, such as the fully-connected,  sub-connected and overlapped subarray structures. By using a cellular infrastructure-to-everything application consisting of both cellular and vehicular communications with 1000 channels and multiple pedestrian users and high-mobility vehicles, the analysis of antenna array structures can be performed. Simulation results show a balanced performance trade-off in terms of spectral efficiency, energy efficiency, and hardware costs between the popular fully-connected structures in VIoT communications with THz massive MIMO. This trend is expected to continue in various areas of THz band-based IoT communications such as  channel modeling and spectrum allocation. In the future, time-varying THz communication issues caused by the high mobility of vehicle should be considered, which could be addressed by predictable AI approaches such as dynamic DRL algorithms}. 
\subsection{Massive Ultra-Reliable and Low-latency Communications}
In the 6G era, it is expected to achieve URLLC for supporting future IoT services through low-latency and reliable connectivity \cite{54}, \cite{55}. \textcolor{black}{For example,  mURLLC is expected to support the timely and highly reliable delivery of massive health data for facilitating remote healthcare, aiming to provide better medical services to patients in the remote areas and also reduce regional imbalance in the health workforce. mURLLC can be deployed in smart factories to automate the mission-critical processes such as automatic manufacturing and remote robotic control. Compared with the traditional wired connections, the use of mURLLC technology allows for optimizing the operational cost with extremely low latency and ultra-high reliability, e.g., the reliability of 99.9999\% with block error rate (BLER) of below $10^{-5}$. Thanks to its outstanding features, mURLLC also well supports transportation systems, by offering timely data sharing of information among vehicles, infrastructures, and pedestrians with high reliabilities, which thus enhances road safety and improves traffic efficiency in vehicular networks. The integration of mURLLC into smart grid is also an active application, aiming for replacing cable/fibre based solutions to carry out the real-time protection and control over the distributed grid lines and stations. As a result, a wide range of  mission-critical services can be realized with mURLLC such as fast fault diagnosis and accurate positioning, reliable fault isolation and system restoration as well as remote decoupling protections. }

Moreover, the recent advances in AI make it an ideal tool to analyze the latency and reliability for enabling mURLLC in 6G-IoT, by offering excellent solutions, e.g., accurate traffic and mobility prediction with deep learning (DL) and fast network control with deep reinforcement learning (DRL) \cite{56}, \cite{406}. The use of AI becomes more significant when the network information is unavailable and the network environment is highly dynamic. For example, the work in \cite{57} exploits DL to a mURLLC-based virtual robotic arm system where deep neural networks (DNNs) are applied to predict accurately arm positions and control robotic arms with a duration of 1 ms. Deep transfer learning is also adopted in the DL architecture to fine-tune the pre-trained DNNs in non-stationary networks for improving the learning efficiency. \textcolor{black}{Moreover, DRL has been used in \cite{58} for optimizing the distributed cooperative sub-channel assignment and transmission power control which aims to provide the strict reliability and latency requirements of URLLC services. By using DRL, each mobile IoT device is able to intelligently make decision on its spectrum access based on its own instantaneous observations, aiming to optimize its sub-channel assignment and transmission power control. Then, a proper QoS-aware reward function is derived to manage the energy efficiency and QoS requirements of all IoT users. Experiments are implemented in a URLLC-based IoT environment with 2000 devices, the minimum data rate requirement is set to 3.5 bps/Hz, and the reliability requirement varies between 99.9\% and 99.99999\%. Implementation results demonstrate a much better energy efficiency by using DRL, compared to other conventional random approaches. A limitation of this approach is the lack of training latency analysis in the DRL running under the latency requirements of URLLC.}

To achieve the latency and reliability for mURLLC, IoT applications should meet several critical requirements. System overheads in term of channel access, user scheduling, and allocation of resources should be minimized. The recent advances in AI open new opportunities for latency optimization in IoT networks, such as optimal IoT user selection and scheduling via DRL approaches, and low-latency resource allocation in intelligent transportation using distributed training techniques via FL. Furthermore, the packet error probability should be minimized to achieve lower-latency data transmission since traditional methods such as hybrid automatic repeat request (HARQ) processing are not appropriate to achieve a low block error rate (BLER) \cite{408}. Additionally, energy-efficient solutions for IoT devices should be designed, aiming to solve the issues of continuously checking awaiting packets on the network that incurs high latency as well. How to implement energy-saving while offering a solution for high frequency of data checks is needed for URLLC-based IoT devices. 

\subsection{Blockchain}
\begin{figure}
	\centering
	\includegraphics[width=0.98\linewidth]{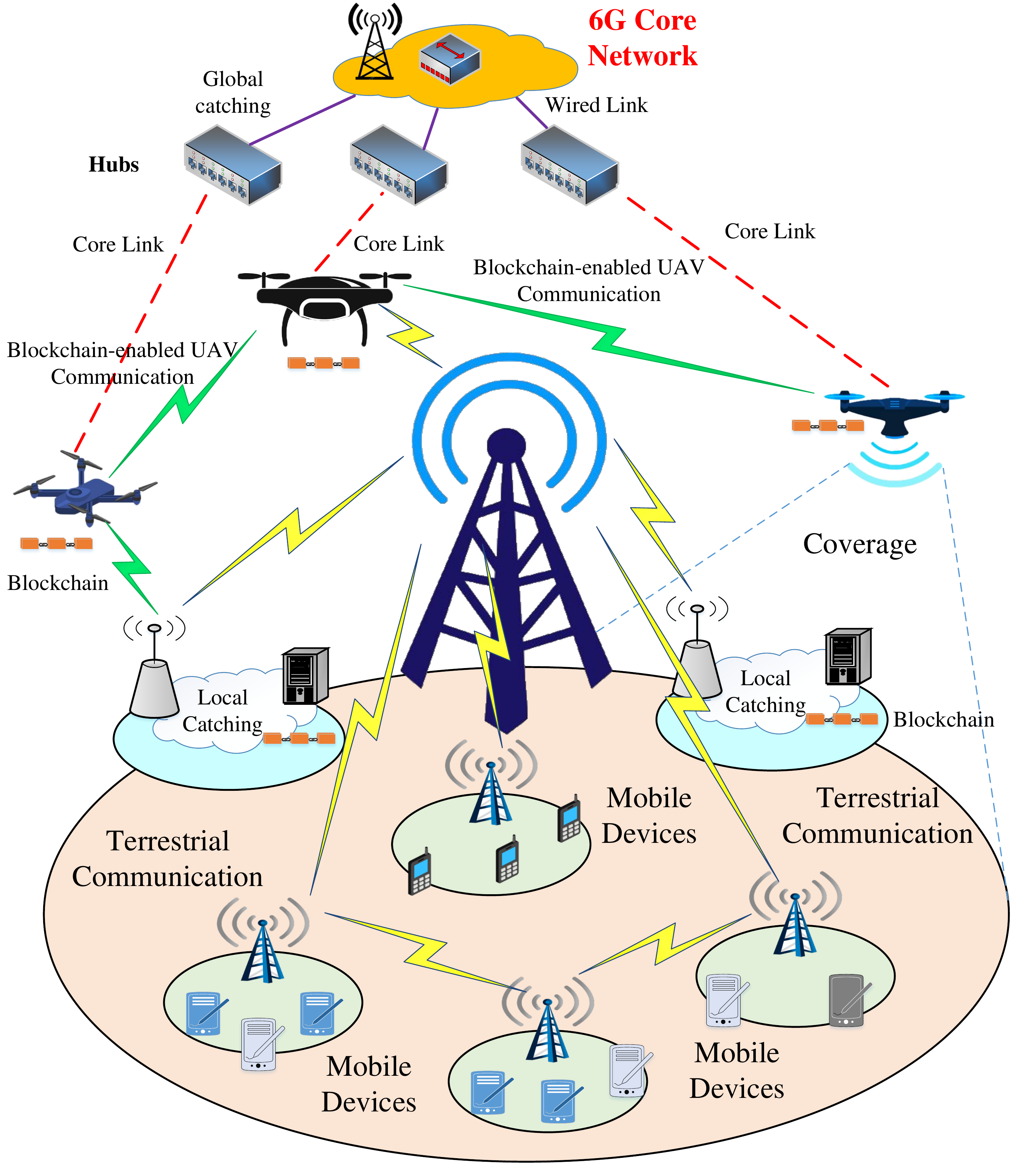}
	\caption{Blockchain for secure 6G UAV networks. }
	\label{Fig:blockchainUAV}
	\vspace{-0.1in}
\end{figure}
In 6G-based IoT networks, how to achieve a high degree of security and privacy is a practical challenge since the fact that 6G systems tend to be distributed and thus suffer from the higher risks of attacks and threats.  Moreover, how to ensure data privacy in the open sharing systems in multi-layer 6G systems such as vehicular data sharing in autonomous driving is a critical issue. Blockchain \cite{59}, as an emerging disruptive technology, is able to offer innovative solutions to effectively deal with such privacy and security challenges in 6G-IoT networks. Conceptually, the blockchain is a decentralized, immutable and transparent database where no any authority is needed to manage the data. This is enabled by a peer-to-peer network topology which allows each entity, e.g., an IoT device, to hold an equal right to control and authorize the data stored in the blockchain. \textcolor{black}{Generally, blockchains can be classified as either a public (permission-less) or a private (permissioned) blockchain. A public blockchain allows anyone to perform transactions and join in the consensus process. The best-known public blockchain applications include Bitcoin and Ethereum. Private blockchains on the other hand are an invitation-only network managed by a central entity, and a participant who wants to join the blockchain network needs to be permissioned via a validation mechanism.} Enabled by its working concept, blockchain boasts a number of desirable characteristics, including decentralization, traceability, trustworthiness, and immutability. These features make blockchain a promising candidate to be integrated into 6G-IoT ecosystems for security and privacy provision. \textcolor{black}{For  secure access control in 6G-based IoT communication environments, a blockchain-based approach is proposed in \cite{61}. The authors exploit the decentralization of immutability features of blockchain to develop a security mechanism for reliable resource access control and user privacy preservation. Specifically, for a secure access control, the states of the virtualized resources are modelled using a Q-learning approach that can learn the resource usage patterns to detect abnormal data access behaviours. In terms of privacy preserving, a joint method of low-latency and memory saving  is used to augment the response success ratio, and to obtain the false positives of the connected users. The benefits of the proposed solution are verified by the high true positives (nearly 94\%), accurate access denial and success ratio (nearly 92\%). However, the mining latency caused by block verification in blockchain has not been considered yet, and the evaluation of data leakage probability is still missing. }

Moreover, a roadmap for the applications of blockchain in 6G-IoT automation is drawn in \cite{62} where blockchain is particularly useful for applied domains such as UAVs, smart grid, and food industry. In fact, blockchain is able to establish secure autonomous systems where UAVs can act as blockchain clients to communicate with ground base stations, aiming to exchange and share data to fulfil their missions such as emergency search or environmental monitoring via a peer-to-peer ledger, as illustrated in Fig.~\ref{Fig:blockchainUAV}. By using blockchain, UAVs, terrestrial users, and network operators can trust the data stored on the ledger with a shared control and tracing right provided over the distributed environment. Another possible IoT application is smart healthcare; the verification of health data 6G-based healthcare systems can be implemented by blockchain and its inherent smart contract technology \cite{63}, where no third-party is required while a high degree of trust is ensured.  \textcolor{black}{Moreover, the application of blockchain in future 6G-IoT networks also might result in costs in terms of latency and energy usage. The mining process, e.g., block verification and information exchange among miners, leads to high network delays and consumes excessive  energy. For example, in the Ethereum blockchain platform, miners need to run Proof-of-Work (PoW)  which is computationally extensive and time-consuming. Further, the repeated information exchange among multiple miners in the block verification also needs large bandwidth resources. Hence, it is essential to take operational costs into account when applying blockchain in future IoT networks. }

To this end, we summarize the fundamental 6G technologies along with their key features and important use cases in IoT networks in Table~\ref{Table:EnablingTech}.
\begin{table*}[h!]
	\centering
	\caption{Taxonomy of fundamental 6G technologies for IoT.}
	\label{Table:EnablingTech}
	\resizebox{\textwidth}{!}{%
		\begin{tabular}{|P{1.8cm}|P{6.5cm}|P{9.5cm}|}
			\hline
			\textbf{Fundamental technology}& 
			 \textbf{Key features }&	
			\textbf{Potential applications in 6G-IoT}
			\\ \hline
			\multirow{11}{*}{\makecell{Edge Intelligence}} &
		\begin{itemize}
			\item	Enabled by the seamless integration of AI, communications, and edge computing where AI functions are deployed at decentralized edge nodes to make intelligence close to the data source where they are generated, e.g., mobile devices.
			\item Privacy-enhanced edge intelligence can be realized via distributed collaborative data training with FL. This learning paradigm allows distributed IoT devices to collaborate with an aggregator to perform AI training while raw data sharing is not needed. 
		\end{itemize} &	
		\begin{itemize}
			\item A self-learning edge intelligence solution is proposed in \cite{32} with a case study in 6G-based vehicular networks \textcolor{black}{where an AI-based GAN is deployed on each vehicle to perform personalized classification of vehicular  latency data without the central processing at a centralized server.}
			\item The work in \cite{33} illustrates the roles of edge intelligence in 6G-IoT via some dominant use cases such as autonomous driving and collaborative robots in smart manufacturing. \textcolor{black}{Compared with \cite{32}, this scheme still has high data communication latency due to the lack of synthetic data production with GANs at edge nodes.}
			\item Edge intelligence in the 6G-IoT context is enhanced with FL \cite{35} for UAV networks where multiple mobile users collaboratively join the FL process with the server deployed at the UAV.
		\end{itemize}
			\\ \hline
			\multirow{10}{*}{\makecell{RISs}} &
			\begin{itemize}
				\item RISs are man-made surfaces of electromagnetic material that are electronically controlled with integrated electronics to enable reconfigurable propagation environments \cite{38}.
				\item The reconfigurability makes RISs as a promising solution to support the wireless system design and optimization to enable smart radio environments and benefit 6G-based IoT applications. 
			\end{itemize} &
			\begin{itemize}
				\item RISs can help establish the interface between the indoor and outdoor entities \cite{39}, aiming to facilitate the access of private households in smart buildings. RISs are also promising to provide cooperative entity layers for avoiding interference and improving spectral efficiencies in the device communications between indoors and outdoors.
				\item RISs are also considered in \cite{42} to support RF sensing for human posture recognition in IoT applications such as surveillance and remote health monitoring. 	
			\end{itemize}
			\\ \hline
			\multirow{12}{*}{\makecell{Space-air-ground\\-underwater \\ Communications}} &
			\begin{itemize}
				\item A cell-free and four-tier large-dimensional communication network for 6G-IoT can be derived, consisting of four tiers: space, air, terrestrial, and underwater \cite{4}.
				\item UAVs can be used as flying stations to provide coverage and connectivity for disaster-stricken areas and supporting public safety networks and emergency situations when URLLC is required. 
			\end{itemize} &
			\begin{itemize}
				\item A multiuser satellite IoT system is considered in \cite{47} based on LEO satellite communications where an MEC server is integrated with the full-duplex access points to create satellite-links for improving communication latency efficiency in mission-critical IoT applications.
				\item	\textcolor{black}{Different from \cite{47}, UAVs are exploited in \cite{48} for supporting different data transmission solutions in IoT data sensing and video streaming, where a new DRL algorithm is integrated to optimize the trajectory and power control of UAVs.} 
			\end{itemize}
			\\ \hline
			
			\multirow{11}{*}{\makecell{THz \\ Communications}} &
			\begin{itemize}
				\item THz communications are envisioned as a driving technology for 6G-IoT which requires 100+ Gbps data rates and 1- millisecond latency \cite{51}. 
				\item The THz spectrum is able to deal with the spectrum scarcity problems in wireless communications and significantly enhance wireless system capacities in 6G-IoT.
			\end{itemize} &
			\begin{itemize}
				\item Recently, THz communication technologies have been used for user localization by exploiting the ultra-wide bandwidths available at THz frequencies \cite{52}.
				\item	\textcolor{black}{THz bands are also exploited for UAVs communications \cite{53}, aiming to analyze the coverage probability of UAV networks. However, unlike \cite{52}, this scheme leverages THz base station density and the strength of THz signals for UAVs' location and trajectory estimation. }
		
			\end{itemize}
			\\ \hline
			
			\multirow{12}{*}{\makecell{mURLLC}} &
			\begin{itemize}
				\item Massive URLLC in 6G-IoT can be realized by mMTC and 5G URLLC integration for enabling mURLLC \cite{55} to provide extremely low latency, extremely reliable connectivity, high availability and scalability 
				\item The recent advances in AI make it an ideal tool to model the latency and reliability for enabling mURLLC in 6G-IoT, by offering excellent solutions, e.g., accurate traffic and mobility prediction with DL, and fast network control with DRL \cite{56}.
			\end{itemize} &
			\begin{itemize}
				\item The work in \cite{57} exploits DL to a mURLLC-based virtual robotic arm system where DNNs are applied to predict accurately arm positions. Deep transfer learning is also adopted in the DL architecture to fine-tune the pre-trained DNNs in non-stationary networks for improving the learning efficiency.
				\item 	\textcolor{black}{ The work in \cite{58} offers a more advanced learning approach, by integrating DNNs with reinforcement learning for optimizing the distributed cooperative sub-channel assignment and transmission power control, which aim to provide the strict reliability and latency requirements of URLLC services.	}
			\end{itemize}
			\\ \hline
			
			\multirow{14}{*}{\makecell{Blockchain}} &
		\begin{itemize}
			\item Although 6G has the capability to provide exceptional service qualities to IoT applications, there exist critical issues in terms of risks of data interoperability, network privacy and security vulnerabilities.
			\item Blockchain is able to offer many innovative solutions to effectively deal with such privacy and security challenges in 6G-IoT networks \cite{59}. Technically, the blockchain is a decentralized, immutable and transparent database where no any authority is needed to manage the data.
		\end{itemize} &
		\begin{itemize}
			\item A blockchain-based approach is proposed in \cite{61} for secure access control and privacy preservation in 6G-based IoT communication environments. 
			\item 	A roadmap for the applications of blockchain in 6G-IoT automation is drawn in \cite{62} where blockchain is particularly useful for applied domains such as UAVs, smart grid, and food industry. For example, blockchain is able to establish secure autonomous systems where UAVs act as blockchain clients to communicate with ground base stations in a secure manner. 
			\item 	\textcolor{black}{The verification of smart healthcare in 6G-based healthcare systems can be implemented by blockchain \cite{63} with the integration of smart contract for self-executing data access evaluation, which has not been considered in \cite{61} and \cite{62}.}
		\end{itemize}
		\\ \hline
		\end{tabular}%
	}
\end{table*}

\section{6G for IoT Applications}
\label{app} 
Enabled by the fundamental technologies as described in the previous section, 6G is envisioned to realize new applications for IoT. In this section, we explore and discuss extensively the emerging applications of 6G in a wide range of important IoT domains, including HIoT, VIoT and Autonomous Driving, UAVs, SIoT, and IIoT. 
\begin{figure}
	\centering
	\includegraphics[width=0.99\linewidth]{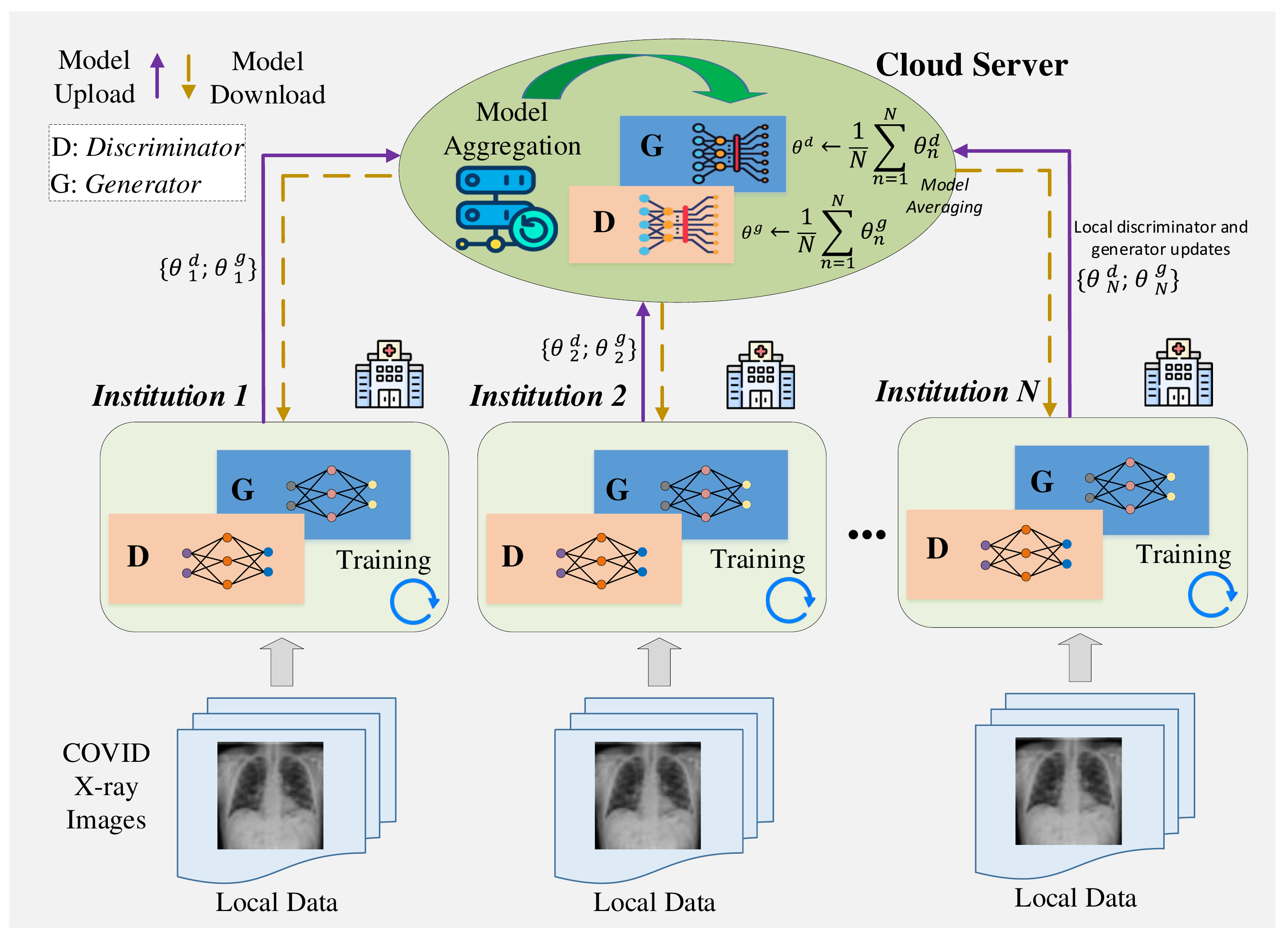}
	\caption{Edge intelligence for COVID-19 data analytics in federated healthcare.  }
	\label{Fig:Covid_EdgeIntelligence}
\end{figure}
\subsection{6G for Healthcare Internet of Things (HIoT)}
The integration of 6G will revolutionize HIoT \cite{64} by using its enabling technologies. The work in \cite{65} discussed the potential use of 6G technologies such as mURLLC and THz communications for supporting extremely low-latency healthcare data transmission and accelerating medical network connections between wearables and remote doctors. \textcolor{black}{In fact, healthcare domains such as remote health monitoring requires low-latency communications (below 1 ms) with the reliability requirement of above 99.999\%   to achieve nearly real-time health provision with a fast and reliable remote diagnosis.} Interestingly, 6G robotics can be applied to implement remote surgery in a fashion that remote doctors can manage the surgery via the robotic systems at a latency of milliseconds and high reliability. Recently, a telesurgery system is also studied in \cite{67} in the context of 6G by using UAVs and blockchain. \textcolor{black}{Given the security risks in existing mobile surgery networks, blockchain is integrated into the robotic system where each robot acts as a data node so that surgical information is stored securely in the database ledger without the need of centralized authority.}  Particularly, smart contracts are also adopted that can provide automatic authentication for health data requests and control over the health data sharing during the surgery \cite{68}. \textcolor{black}{To solve the slow healthcare response rates, UAVs are employed as relays  to  transport light-weight healthcare items such as medicines and surgical tools among hospitals in emergent situations, which helps avoid road-traffic congestions and thus mitigates data exchange latency.} Meanwhile, to achieve future requirements in terms of ultra-high data rates of medical data communications, the mURLLC technology is an ideal solution by using the THz bands in 6G-based healthcare networks \cite{69}. Accordingly, nano-devices, implants, and on-body sensors can communicate and transmit data in real-time with extremely high reliability and availability to edge devices or cloud centres for short- and long-term medical analysis. In particular, mURLLC also plays a key role in hospital-based telestration where doctors can monitor and manage the surgery procedure remotely using real-time video streaming from medical robotics and assistant devices interconnected by 6G core networks \cite{70}. \textcolor{black}{Moreover, 6G-based URLLC has been exploited to facilitate connected ambulance in future healthcare, by allowing real-time video streaming with high color resolution for reliable diagnosis to clinicians and paramedical staff from the hospital at moderately high speeds (up to 100 km/h) \cite{71}.} \textcolor{black}{For example, electroencephalogram data from clinical examination can be conducted on-board and through URLLC-based real-time teleconsultation, hospital doctors are able to provide urgent indications to paramedical staff in the ambulance.} \textcolor{black}{In this scenario, a very small survival time must be ensured (below 2 ms) although the ambulance can run at high speeds.} 

To realize intelligent 6G-based healthcare, AI can be exploited for data learning and analytics. The study in \cite{72} uses various machine learning (ML) techniques such as Bayesian classifier, logistic regression, and decision tree to analyze historical health records of stroke out-patients collected from wearable sensors in heathcare-based 6G heterogeneous networks. To accelerate the stroke care for patients, an uplink radio resource allocation optimization solution is integrated where the assigned resources are proportional to the stroke likelihood of patients. Another ML-based solution for intelligent 6G-healthcare networks is suggested in \cite{74}, where edge-cloud computing is adopted to provide low-latency health data analytics for healthcare services such as diagnosis, disease prediction, and intelligent decision making tasks for physical medicine and rehabilitation. In this context, ML is also useful to optimize mobility management processes in 6G-based health networks, by taking data rates, traffic flows, data processing delays, and bandwidth resource allocation into account. Implementation results show a good trade-off between time and energy efficiency by using ML techniques while effectively managing and monitoring the mobility of the IoT driven devices in 6G-empowered industrial applications including healthcare services. Recently, COVID-19 has spread rapidly across the globe and become a major health concern of many countries. Wireless communication technologies such as URLLC, edge intelligence, and cloud computing have been applied to combat the COVID-19 pandemic in different ways \cite{75}. For example, high-speed live video conferencing based on URLLC enables healthcare professionals to discuss with patients in a low-latency and reliable manner for timely COVID-19 outbreak analysis. COVID-19 data can be processed using data-driven edge intelligence techniques by integrating edge computing and AI for accurate and fast disease diagnosis \cite{76}. \textcolor{black}{Fig.~\ref{Fig:Covid_EdgeIntelligence} illustrates a case study of using edge intelligence for COVID-19 analysis, by applying FL at the network edge. \textcolor{black}{Each edge server located at a local hospital institution is equipped with a local GAN consisting of a discriminator and a generator based on CNNs to learn the COVID-19 data distribution using its own local image dataset.} Then, the local GANs synchronize and exchange the learned model parameters for aggregation at a cloud server, which then returns a new version of the global model to all institutions for the next training round.} This process is repeated until a desired accuracy is achieved, aiming to generate realistic COVID-19 images for the detection of COVID-19. The use of edge intelligence thus offers unique benefits to COVID-19 analytics, including privacy protection and large-scale data processing capabilities. 
\begin{figure*}
	\centering
	\includegraphics[width=0.99\linewidth]{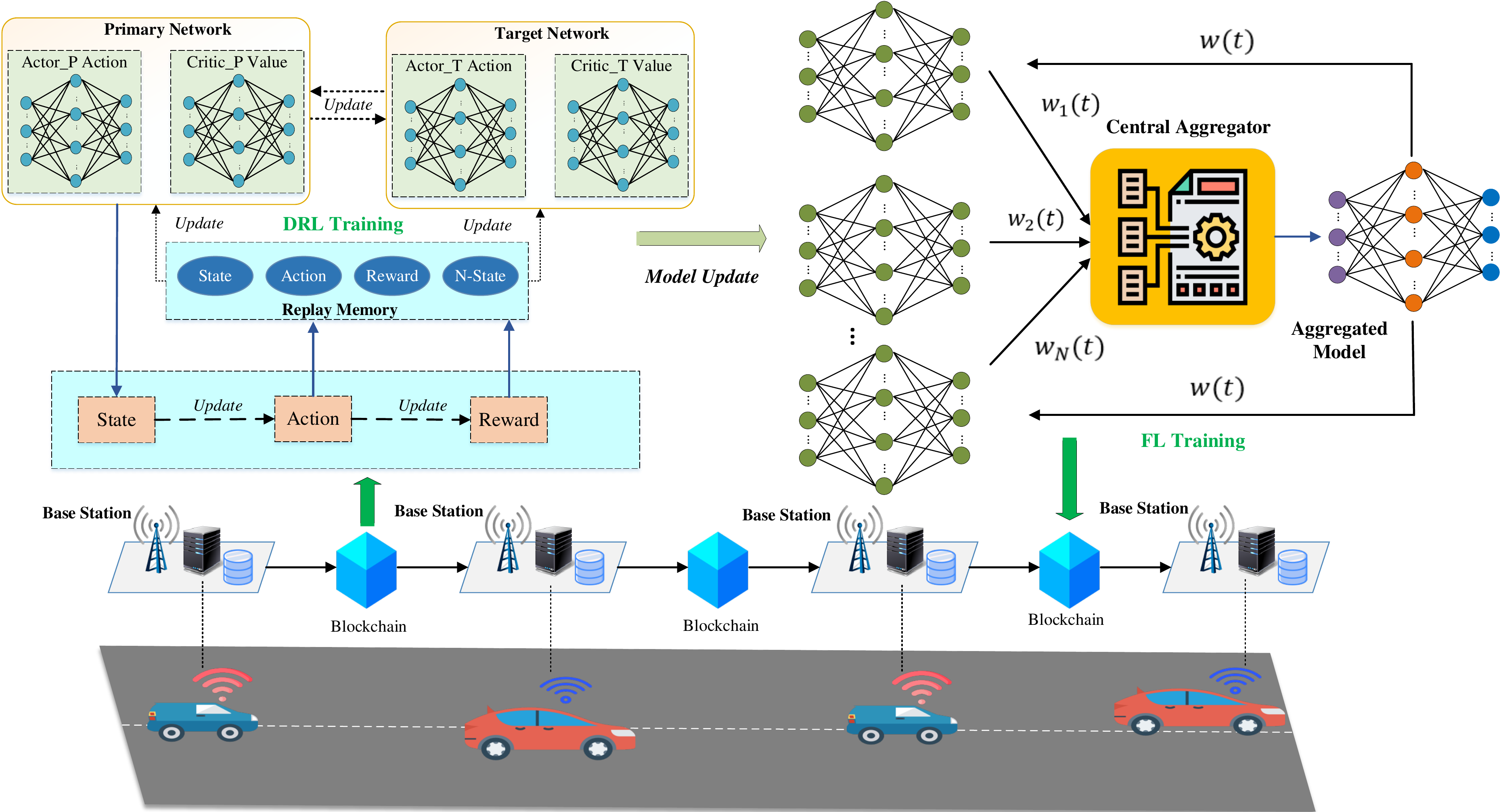}
	\caption{DL for V2V network management.  }
	\label{Fig:Dl-Vehicular}
	\vspace{-0.1in}
\end{figure*}
\subsection{6G for Vehicular Internet of Things (VIoT) and and Autonomous Driving}
The advances in 6G technologies have greatly transformed vehicular Internet of Things (VIoT) networks and thus revolutionized intelligent transportation systems (ITSs). The work in \cite{77} exploits mMTCs to enable vehicle-to-everything (V2X) connectivity for the transmission of short vehicular information payloads by a high number of vehicles without human interaction in 6G-based VIoT networks. To do so, the idiosyncrasies of V2X are taken to strike a trade-off between scalability, reliability, and latency via a vehicle discovery approach in which a discovery entity located at the base station collects information about the proximity of the vehicles. Accordingly, to optimize the discovery scheme, signature properties such as time slots and hash functions are tuned, aiming to minimize the false-positive probability to schedule the radio resources for V2X data communications under the available spectrum budget. In  future 6G-based VIoT, data rate prediction is a challenging task due to the complex interdependency between factors such as mobility, channels, and networking. ML can be an efficient approach to mimic the possible behavior of network-assisted throughput prediction in future 6G vehicular networks \cite{78} by learning the historic network load information based on control channel analytics. \textcolor{black}{ To unleash the potential of vehicular intelligence in VIoT, edge intelligence functions with ML are integrated into road side units (RUSs) that are responsible for performing the estimation of traffic volume and weather forecast based on the aggregation of local observations from vehicles \cite{79}.} To further enhance the scalability of vehicular systems in the context of 6G, a distributed estimation approach is proposed, by allowing for local estimation at distributed vehicles through wireless data exchange with neighbouring vehicles within the communication range. 

The role of DL in providing intelligence for 6G-based VIoT is also examined in \cite{80} by leveraging its high-dimensional generalization ability to model vehicular communication channels and support networking management, such as optimal resource allocation using DRL algorithms \cite{81}, as illustrated in Fig.~\ref{Fig:Dl-Vehicular}. In this context, how to build trust for data learning and reasoning is highly important as the data training is normally treated as a process in a black box where the input and output are known only. Therefore, a trust broker entity is proposed for 6G-VIoT networks, which is able to provide reasoning for learning actions at the DL controller, e.g., the base station. Thus, the learning process can be interpreted so that business stakeholders can understand the data training process, e.g., which data features cause which decisions. Vehicular intelligence is also the focus in \cite{82} where DL techniques are adopted to autonomously schedule the data transmission. \textcolor{black}{This is enabled by using three approaches, supervised learning for data rate estimation, unsupervised learning for recognizing geospatially-dependent uncertainties of the estimation model, and reinforcement learning (RL) for autonomously coordinating data transmissions based on anticipated resource efficiency.} This combined scheme is promising to achieve multi-objective optimization in VIoT, from resource allocation to data rate maximization. In the future, it is useful to develop cooperative DL approaches like FL since 6G-VIoT networks are expected to be highly scalable and distributed. In addition to network management, \textcolor{black}{DL-based intelligent software is deployed at the vehicular data controller that is able to provide useful solutions for security protection in 6G-based VIoT networks \cite{83}.} \textcolor{black}{A new weight-based ensemble DL approach is suggested to detect intrusion and attack risks in vehicular communications. Support vector machine is firstly used to map data to a high-dimensional space through a kernel function and build the optimal classification hyperplane to extract the sample classes. K-neighborhood is then applied to determine the categories of extracted samples, which is then classified by a decision tree. Numerical results demonstrate the high performances of DL in terms of high attack detection accuracy (an increase of 5\%-20\%, compared to non-learning approaches), which thus helps improve the reliability and security of vehicle networks in 6G. }

In the future vehicular networks, autonomous driving (AV) will play a significant role in improving transportation quality, road safety, and vehicular energy efficiency. 6G technologies are expected to provide exciting opportunities to meet the stringent service requirements of AV applications for reliability and high-speed communications \cite{84}. To fully realize AV in the 6G, it is important to investigate the communication performances of the V2V networks since each vehicle is regarded as an entity with full control and recognition in the interconnected vehicular system. In this context, cooperative driving is enabled through information sharing and driving coordination among vehicles, where  DL can come as a natural solution to perform fast prediction of the V2V communication performance bounds for intelligent control of inter-vehicle distances.  \textcolor{black}{Edge intelligence is also of paramount importance for providing intelligent functions at the network edge, e.g., RSUs, for controlling the AV system \cite{85}.} For example, AV controllers can be located at edge servers embedded with DL processors to train vehicular data for implementing autonomous driving decision making and high-definition mapping for navigation. FL can also be exploited to provide cooperative learning and perform federated vehicular communications among vehicles and edge servers, while preserving user privacy and reducing network overheads caused by raw data sharing \cite{86}.

\subsection{6G for Unmanned Aerial Vehicles (UAVs)}
Enabled by the emerging wireless technologies, many research efforts have been put into exploring the applications of 6G-UAV networks. The work in \cite{87} considers a cell-free UAV network for 6G-based wide-area IoT with focus on a UAV flight process-oriented optimization. This can be done by formulating a data transmission efficiency maximization problem that can be solved by taking large-scale channel state information, on-board energy, and interference temperature constraints into account. The proposed approach is also promising to identify the cell-free coverage patterns to support massive access for wide-area IoT devices in the future 6G era. The authors in \cite{88} propose a UAV-supported clustered non-orthogonal multiple access (C-NOMA) scheme \cite{206}  for supporting wireless powered communications in 6G-enabled IoT networks. Given the popularity of cluster IoT terminals, a terminal clustering strategy is adopted based on intra-cluster NOMA communications which allows UAV to transmit radio signals to IoT terminals. A synergetic optimization solution for UAV trajectory planning and subslot allocation is derived by portioning the downlink energy transfer subslot and uplink information transmission subslot. This aims to maximize the achievable sum rates of all IoT terminals which are confirmed via numerical simulations. Unlike the work in \cite{88}, the project in \cite{89} pays attention to characterize the UAV-to-ground channel with arbitrary three-dimensional (3D) UAV trajectories for UAV-based 6G networks. To do so, a 3D nonstationary geometry-based stochastic model is derived by using the multiple-input multiple-output (MIMO) channel configuration, with respect to distinctions between AV altitudes, spatial consistency, and 3D arbitrary UAV movement trajectories. \textcolor{black}{Meanwhile, a collaborative multi-UAV trajectory optimization and resource scheduling framework is studied in \cite{90} in a 6G-IoT network, where multiple UAVs are used as flying base stations to transfer energy to multiple terrestrial IoT users.} The design is focused on the association between UAVs and users, by implementing a user association solution to select the most appropriate user to upload data to a specific UAV. \textcolor{black}{Accordingly, the ultimate objective is to optimize the average achievable rate among all IoT users, with respect to the UAV trajectory, sub-slot duration, and user transmit power. A joint algorithm based on the relaxation and successive convex optimization methods is derived to solve the proposed problem, showing a better achievable rate compared to existing schemes.} \textcolor{black}{Similarly, the research in \cite{91} concentrates on optimizing the transmission rates of UAVs, where UAVs act as mobile relays in NOMA-based cognitive 6G-IoT networks.} A flexible approach is adopted that allows for optimally selecting relays to provide higher transmission rates under fixed power.

In addition, AI techniques have been used to provide intelligent solutions for 6G-based UAV networks. In fact, UAVs can provide wireless communication services, edge computing services, and edge caching services when empowered AI-based solutions \cite{92}. For example, to control the UAV mobility and mission scheduling for the UAV trajectory planning, AI techniques like ML can be exploited to predict the future demands of users and service areas based on historical datasets of movement trajectories and user requests. This not only adjusts optimally the UAV trajectories to save transmit power but also enhances the quality of users' experience. Moreover, AI approaches are very useful to build proactive edge data caching in UAV-based IoT networks, based on the data training and prediction capabilities through learning and feedback processes, e.g., data caching with DRL \cite{93}. Very recently, FL has been applied in \cite{94} to provide privacy-preserved intelligence for UAV-based 6G networks. \textcolor{black}{Here, each UAV runs a DL model and exchanges learned parameters with an MEC server for aggregation.} To accommodate the federated data training in the UAV network with limited batteries and bandwidth spectrum, a resource allocation problem is considered that is then solved by a DRL algorithm. \textcolor{black}{In the future, to facilitate operations of UAVs in 6G-IoT networks, regulations should be put into place  to provide guidelines for the deployment of UAVs in IoT systems, to ensure safety and privacy  \cite{405}. Moreover, local licensing regulations are also important, especially when countries are still in the process of defining rules for spectrum access rights to UAV manufacturers. Regulatory authorities should collaborate to address critical issues in UAV integration into the existing IoT networks, from pricing strategies to network deployment choices, interference protection and aerial service coordination. }

\subsection{6G for Satellite Internet of Things (SIoT)}
In the 6G era, it is highly essential to integrate satellite communications into current wireless networks for massive IoT coverage which gives birth a new domain  called SIoT \cite{95},  \cite{407}. Conceptually, satellites consist of three main network tiers, namely LEO, MEO, and GEO to offer global services to the terrestrial IoT users. \textcolor{black}{Compared to MEO and LEO, the LEO system has been received priorities in 5G-network generation research due to its lower orbit height and useful features to support IoT connectivity, such as shorter transmission delay, smaller path loss.  However, in the 6G era, thanks to the advanced satellite technologies, multiple satellites can be deployed at dozens of orbits above the earth, LEO systems can thus authentically realize the global coverage and more efficiency by frequency reuse. Moreover, it is envisioned to establish inter-satellite links to enable inter-satellite communications based on THz bands, which can accommodate more satellites and achieve higher link performances due to its much wider bandwidth, compared to existing spectrum resources in mmWave communication and optical communication counterparts in the 5G era \cite{95}.} As a case study in SIoT, the work in \cite{96} focuses on research of a LEO satellite network that can support the navigation of UAV trajectories for IoT data collection missions. A two-mode communication model is derived, i.e., the UAV carry-store mode and the satellite network relay mode for IoT data transmission. In this context, a UAV energy cost optimization problem is formulated that is then solved by a column generation based algorithm. \textcolor{black}{To provide further insights into SIoT, the authors in \cite{97} build a comprehensive model along with highlighting technical discussions and challenges; this includes satellite base stations and a network of terrestrial users where satellite communications can be realized via relays located at terrestrial base stations.} Satellite communications can be integrated with the RIS technology to enhance transmit power consumption at the satellites, as shown in \cite{98}; thus, a RIS-assisted LEO satellite framework is created. RIS units can be set up on the rectenna arrays to support IoT data broadcasting and beamforming based on transmit features such as carrier frequencies. In line with the energy discussion, an energy-aware massive random access scheme for satellite communications in 6G-enabled global SIoT is considered in \cite{100}. \textcolor{black}{In reality, to communicate to the base station via the uplinks, each IoT device needs to create a random access process, by selecting an available preamble from the provided preamble set for data transmission. Accordingly, an enhanced preamble sequence scheme is proposed to perform one-step fractional timing advance estimation to avoid additional signalling overhead and energy costs. Simulation results confirm a high performance in terms of an increase of up to 15\% in timing estimation accuracy compared to existing approaches. }

\begin{figure*}
	\centering
	\includegraphics[width=0.8\linewidth]{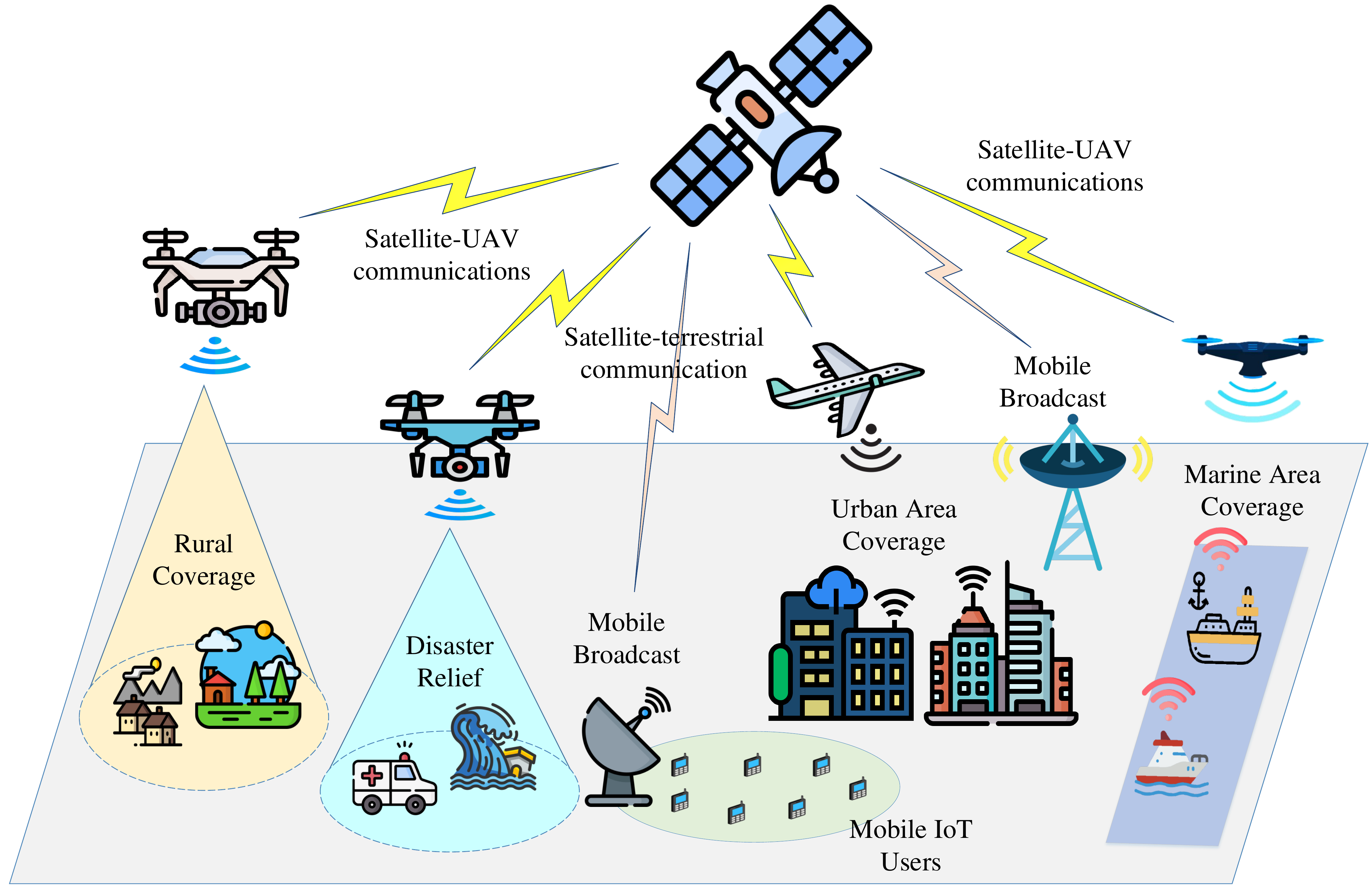}
	\caption{The 6G-based SIoT network and use cases.   }
	\label{Fig:6G_SIoT}
\end{figure*}

In addition to energy management, spectrum sharing is another critical issue in the satellite communications for 6G-based IoT networks \cite{99}. The combination of NOMA and cognitive radio technologies can help to overcome the spectrum scarcity. This is based on that fact that NOMA is able to enhance the spectrum efficiency by enabling users to transmit on the same carrier and classifying users by different power levels. Furthermore, the cognitive radio facilitates dynamic spectrum sharing for efficient spectrum usage. Towards a trade-off between system flexibility, network performance, latency and energy management in 6G-based SIoT networks, a holistic architecture is provided in \cite{101} with different useful applied domains, as illustrated in Fig.~\ref{Fig:6G_SIoT}. For instance, SIoT is flexible to offer remote connectivity where terrestrial networks are overloaded or not possible, like seas and deserts. Further, it is possible to provide data services, such as data offloading and caching to support service delivery for terrestrial base stations with billions of connected users. \textcolor{black}{SIoT platforms can enable energy sustainability via the use of aerial IoT devices such as UAVs and balloons, with renewable sources from the space that may be not available at the ground-based stations \cite{102}.}
\subsection{6G for Industrial Internet of Things (IIoT) }

Recently, the roles of 6G have been investigated in the IIoT domain.  For example, ML approaches have been applied to provide intelligence for 6G-based IIoT networks. The study in \cite{103} investigates the potential of ML-based CNNs in optimizing resource allocation in massive IIoT systems with 6G through a multi-agent system. Due to the limited resource of IIoT sensors, the deployment of sensor nodes is often implemented randomly, introducing unnecessary energy costs. CNNs are useful to perform intelligent sensor clustering via data mining and predictions based on neural backpropagation and interaction in the training process with historical datasets. Numerical simulation verifies a better resource allocation with lower power consumption and reduced complexity. \textcolor{black}{Meanwhile, transfer learning is adopted in \cite{104} to coordinate the data distribution and transmission in a blockchain-based 6G IIoT network, where blockchain is integrated into edge nodes and a cloud server for building a secure data sharing platform.} An integrated fog-cloud computing model is integrated to handle the data analytics and blockchain data, by taking the block size, CPU and memory usage, and network delays into account. \textcolor{black}{The reasoning ability of ML is exploited in \cite{105} for dealing with fault detection and fault prediction problems in 6G-based IIoT systems with industrial sensors and actuators. To this end, a long short-term memory network with a DNN is integrated to perform online fault prediction in a mini manufacturing system, aiming to optimize plant operations and accelerate the production process.  Online learning enabled by data-driven ML is also helpful to 6G-based IIoT network operation interpretation thanks to the reasoning ability of the deep neural structures \cite{106}.} This is particularly important for future 6G-IIoT applications where data volumes are extremely large and its architecture is highly complex which can be explained via data learning.

\textcolor{black}{Intelligent agriculture can be realized where AI techniques would be very useful to optimize agricultural processes, while blockchain is able to enable secure production and logistics, e.g., secure package delivery from the farms to the supermarkets via immutable block ledger \cite{60}. Further, space-terrestrial communications with UAVs has the great potential to advancing precision agriculture, by allowing for aerial-based soil measurement via the sensing capabilities of UAVs over the large-scale coverage \cite{201}. UAVs can be also exploited to support crop imaging from a low altitude, aiming to provide a holistic view on the farm for automatic management of agriculture production. Beyond these sectors, mining is another industrial domain that can be also received much benefits from communication advances \cite{202}, through service management solutions offered by 6G technologies, e.g., massive URLLC-enabled robotics for low-latency and automatic mineral manufacturing, and AI for mineral predictive production control and pricing prediction. }

Although 6G brings unprecedented benefits to IIoT, security and privacy represent significant challenges to be addressed. Blockchain with its decentralization and immutability has emerged as a promising solution to provide security and trust for 6G-based IIoT networks. A secure data aggregation approach for 6G-IIoT systems is considered in \cite{107} based on blockchain. This can build decentralized databases stored over the IIoT devices without the need for an authority. Instead, it relies on the consensus of all participating nodes that can verify and authenticate the task assignment and the task data. In an effort to further enhance privacy protection, a sensitive tasks decomposition and task receivers grouping method is integrated; hence, the direct disclosure of privacy in sensitive tasks. \textcolor{black}{Another blockchain-based solution for security enhancement in 6G-IIoT networks is also suggested in \cite{108} along with FL that manages the industrial data learning process via the collaboration of end users and a base station. In this context, to reduce the unreliable and long-distance communications between end users and edge servers, digital twins are exploited to bridge the physical IIoT systems with the digital world for robust FL training.}

In summary, we list the 6G-IoT applications in Table~\ref{Table:FLchain_Applications} to give better insights into the technical aspects of each reference work.

\begin{table*}[]
	\centering
	\caption{Taxonomy of 6G-IoT applications.  }
	{\color{black}
		\label{Table:FLchain_Applications}
		\begin{tabular}{|p{1.6cm}|p{0.6cm}|P{3cm}|P{2cm}|P{9cm}|}
			\hline
			\textbf{Applications}& 	
			\textbf{Ref.} &	
			\textbf{Use Case}&	
			\textbf{Applied 6G technology}&
			\textbf{Key Contributions} 
			\\
			\hline
			
			\multirow{16}{*}{\makecell{6G-based HIoT}} & 
			\cite{65} &	Remote health monitoring&	mURLLC and THz communications&	A solution for remote health monitoring with extremely low-latency data transmission and network connections between wearables and remote doctors.
			\\ \cline{2-5}&
			\cite{67}&	Remote surgery&	UAVs and blockchain &\textcolor{black}{A more comprehensive model using the UAVs and blockchain integration that can provide both fast and secure features for surgery processes.}
			\\ \cline{2-5}&
			\cite{69}&	Healthcare data communications&	mURLLC&	An improved healthcare communication approach using mURLLC for communication between nano-devices, implants, and on-body sensors and edge/cloud servers. 
			\\ \cline{2-5}&
			\cite{71}&	Connected ambulance&	URLLC&	A scheme using 6G-based URLLC to facilitate connected ambulance in future healthcare, by allowing real-time video streaming with high color resolution for reliable diagnosis. 
			\\ \cline{2-5}&
			\cite{72}&	Health data analytics and resource allocation optimization&	ML&	A ML-based framework to analyze historical health records of stroke out-patients collected from wearable sensors and radio resource allocation optimization in heathcare-based 6G heterogeneous networks. 
			\\ \cline{2-5}&
			\cite{74}&	Mobility management&	ML&	A solution using ML for ML to optimize mobility management processes in 6G-based health networks. 
			\\ \cline{2-5}&
			\cite{76}&	COVID-19 data analytics&	Edge intelligence&	\textcolor{black}{Edge intelligence-based solutions for COVID-19 data learning and diagnosis in hospital networks. Compared with other works \cite{72, 74}, this scheme involves high training latency due to highly deep CNN structure.} 
			
			\\ \cline{2-5}
			\hline
			\multirow{13}{*}{\makecell{6G-based VIoT}} & 
			\cite{77}&	Enabling V2X connectivity&	mMTC&	 An approach using mMTC to enable V2X connectivity for the transmission of short vehicular information payloads in 6G-based VIoT networks.
			\\ \cline{2-5}&
			\cite{79}&	Vehicular network estimation&	ML&	An ML-based scheme to perform estimation of traffic volume and weather forecast based on the aggregation of local observations from vehicles at RUSs.
			\\ \cline{2-5}&
			\cite{80}&	Vehicular communication modelling &	DL&	\textcolor{black}{Unlike \cite{79}, this scheme relies on the high-dimensional generalization ability of DL to model vehicular communications in 6G-based VIoT.}	
			\\ \cline{2-5}&
			\cite{82}&	Vehicular intelligence&	DL&	An approach for enabling vehicular intelligence where DL techniques are adopted to autonomously schedule the data transmission.
			\\ \cline{2-5}&	
			\cite{84}&	Autonomous driving&	DL&	A scheme for cooperative driving enabled by DL-based fast prediction of the V2V communication performance bounds towards intelligent control of inter-vehicle distances.
			\\ \cline{2-5}&	
			\cite{85}&	Autonomous driving&	DL&	\textcolor{black}{A model using DL-based edge intelligence that is mainly applied to implement autonomous driving decision making and high-definition mapping for vehicular navigation.}  
			\\ \cline{2-5}
			\hline
			\multirow{13}{*}{\makecell{6G-based UAV}} & 
			\cite{87}&	UAV flight process-oriented optimization&	UAV&	A cell-free UAV network for 6G-based wide-area IoT with focus on a UAV flight process-oriented optimization. 
			\\ \cline{2-5}&
			\cite{88}&	UAV wireless powered communications&	UAV& \textcolor{black}{A UAV-based scheme for supporting wireless powered communications in 6G-enabled IoT networks. Compared with \cite{87}, UAV communications are further supported by the NOMA technology, which allows UAV to transmit radio signals to IoT terminals.}
			\\ \cline{2-5}&
			\cite{90}&	Multi-UAV trajectory optimization and resource scheduling&	UAV&	A collaborative multi-UAV trajectory optimization and resource scheduling framework in a 6G-IoT network.
			\\ \cline{2-5}&
			\cite{92}&	Intelligent UAV platforms&	ML& UAVs platforms  can provide wireless communication services, edge computing services, and edge caching services with ML techniques. 
			\\ \cline{2-5}&	
			\cite{94}&	Federated UAV communications&	FL&	\textcolor{black}{As a further design advancement, a privacy-enhanced ML approach is employed via FL to provide reliable intelligence for UAV-based 6G networks.}
			\\ \cline{2-5}
			\hline
			\multirow{12}{*}{\makecell{6G-based SIoT}} & 
			\cite{96}&	Satellite IoT data collection&	UAV&	A LEO satellite network to support in the navigation of UAV trajectories for IoT data collection missions.
			\\ \cline{2-5}&
			\cite{98} &	RIS-based satellite communications&	RIS&	A model for RIS-based satellite communications to enhance transmit power consumption at the satellites, aiming to form a RIS-assisted LEO satellite framework.
			\\ \cline{2-5}&
			\cite{100}&	Energy-aware massive random access &	Satellite communications&	An energy-aware massive random access scheme for satellite communications in 6G-enabled global SIoT.  
			\\ \cline{2-5}&
			\cite{99}&	Satellite  spectrum sharing&	Satellite communications&	A study for spectrum sharing in satellite communications for 6G-based IoT networks. \textcolor{black}{Moreover, this scheme is integrated with NOMA to further enhance the spectrum efficiency.}
			\\ \cline{2-5}
			\hline
			\multirow{8}{*}{\makecell{6G-based IIoT}} & 
			\cite{103}&	Optimal industrial resource allocation&	ML&	An ML-based CNNs approach for optimizing resource allocation in massive IIoT systems with 6G through a multi-agent system.
			\\ \cline{2-5}&
			\cite{104}&	Transmission latency minimization &	ML&	\textcolor{black}{A transfer learning-based method to coordinate the intelligent data distribution and sharing. Particularly, this scheme offers security features by using an immutable data control solution with blockchain. }
			\\ \cline{2-5}&
			\cite{107}&	Secure industrial data aggregation&	Blockchain&	A secure data aggregation approach for 6G-IIoT systems based on blockchain that can build decentralized databases stored over the IIoT devices. 
			\\ \cline{2-5}&
			\cite{108}&	Security enhancement in IIoT&	Blockchain and FL&	A blockchain-based solution for security enhancement in 6G-IIoT networks. \textcolor{black}{Compared with related works \cite{104, 107}, this scheme is integrated with FL that further improves data privacy without sharing of raw data.}
			\\ \cline{2-5}
			\hline
	\end{tabular}}
\end{table*}
\section{Research Challenges and Future Directions}
\label{cha} 
In this section, we highlight several interesting research challenges and point out possible future directions in 6G-IoT.
\subsection{Security and Privacy in 6G-IoT}
The use of 6G technologies will revolutionize the IoT networks and services, with many network features such as high reliability, ultra-low latency, and massive wireless coverage. \textcolor{black}{ However, the integration of 6G into IoT networks may be vulnerable to threats related to wireless interface attacks such as unauthorized access to data at computing units/servers, threats to integrity in the access network infrastructure, and denial of service (DoS) to software and data centres \cite{411}. For example, the diversity of IoT devices and access mechanisms as well as massive device connectivity in large-scale IoT access networks brings new security challenges as handovers between different access technologies increase the risk of attacks. The growth of connections between devices and computing nodes at the network edge also increases the security and privacy threats, where eavesdropping attacks, hijacking attacks, spoofing attacks and DoS attacks may occur in data communications and data management centres. Additionally, to realize intelligent 6G-IoT networks, AI functions can be deployed at distributed edge nodes, where the data training can be manipulated in a spectrum access system by inserting fake signals or modified parameters. Thus, a malicious attack can illegally take advantage of a large portion of spectrum by denying the spectrum to other users. Attackers can also exploit the distributed data training nature and the dependencies on edge computing to launch different attacks such as malicious data injection, data poisoning or spoofing that adversely affect the training outputs of AI functions in intelligent 6G-IoT systems. } Also, edge intelligence can face security vulnerabilities due to the distribution of AI functions at the network edge, where attacks can deploy data breaches or modifications while the management of remote 6G core network controllers is limited \cite{109}. Moreover, the deployment of satellite-UAV-IoT communications over the untrusted environments in space can be hindered by data privacy leakage caused by third-parties and adversaries during the data exchange and transmission between satellite base stations, UAVs, and terrestrial IoT users.

Therefore, the risk mitigation  must be considered to ensure high degrees of security and privacy for 6G-IoT. For example, perturbation techniques \cite{111} such as differential privacy or dummy  can be used to protect training datasets against data breaches in the edge intelligence-based 6G-IoT networks, by constructing composition theorems with complex mathematical solutions \cite{203}. \textcolor{black}{As an example, differential privacy is applied in \cite{112} by inserting artificial noise (e.g., Gaussian noise) to the gradients of NN layers to preserve training data and hidden personal information against external threats while guaranteeing convergence. \textcolor{black}{A novel privacy-preserving data aggregation solution is also integrated under fog computing architecture to satisfy $\epsilon$ differential privacy in the fashion that the aggregation results are close to the actual results while adversaries cannot extract the ground truth in the exchanged gradients.} A Reference Energy Disaggregation dataset including specific information about the electricity consumption of households and a healthcare dataset including more than 1 million medical records are employed for simulations. Implementation results demonstrate that differential privacy helps achieve an up to 6\% higher data protection degree, compared to traditional Laplace differential privacy approaches, under various privacy budget settings.} However, the application of differential privacy also comes with the cost of training quality degradation. In future work, it is suggested to develop accuracy-aware differential privacy designs to strike the trade-off between training quality and privacy protection. Moreover, blockchain is a promising solution to build trust and establish secure decentralized communications for UAV-IoT networks \cite{113}. \textcolor{black}{Each UAV can act as a blockchain node to perform decentralized data sharing and communications, where lightweight mining mechanisms should be also designed for low-latency data consensus with respect to the resource constraints of flying devices like UAVs.} This blockchain technique becomes more important in the 6G era since IoT networks tend to be decentralized and deployed over the large scale that is well suitable for the decentralization feature of blockchain. 
\subsection{Energy Efficiency in 6G-IoT}
In future 6G-based IoT networks, how to achieve high energy efficiency is a major concern. The data communications and service delivery services, e.g., vehicular data sharing in autonomous driving, packet delivery in the space with UAV communications, requires significant energy resources to ensure the network operations. Besides, each base station in wireless cellular networks normally consumes 2.5 kW to 4 kW \cite{114}, which means that the deployment of massive 6G-IoT networks with thousands of stations results in enormous energy consumption that also increases carbon emissions. Designing energy-efficient communication protocols via optimization is desired to realize green 6G-IoT networks. For example, the work in \cite{115} jointly optimizes the QoS and energy consumption in 6G-based smart automation systems. This can be done by implementing a 6G-driven multimedia data structure model with respect to QoS parameters, such as packet loss ratio and average transfer delay during energy-efficient multimedia transmission. 

Energy harvesting techniques to exploit the renewable energy resources would be very useful to build green 6G-IoT systems, e.g., IoT devices can harvest power from ambient environments, such as wind power, solar power, vibration power, and thermal power to serve their communications and computing services \cite{116}. \textcolor{black}{\textcolor{black}{For example, a solar energy harvesting solution is considered in \cite{301} for HIoT networks, where implantable sensors can harvest solar power from natural sunlight for serving the sensory data transmissions via a Bluetooth low energy module in a transparent silicon housing for HIoT-based healthcare monitoring.} Experiments are conducted via a wireless implantable sensor prototype with a solar panel and access point working over a 10 minutes operation cycle, showing a stable energy harvesting while the lifetime of the HIoT-based healthcare system is significantly improved.} The future researchers should further investigate the energy efficiency issues in 6G-IoT networks at higher altitudes, such as satellite networks with flying devices and base stations, where space communication technologies are adopted and energy harvesting is dependent on device trajectories and ambient environments.
\subsection{Hardware Constraints of IoT Devices}
The hardware constraint of IoT devices is another possible challenge in communications and computations 6G-based IoT networks. For example, in intelligent 6G-based healthcare, wearable sensors and mobile devices should be able to simultaneously run AI functions to achieve edge intelligence and implement data transmission with URLLC. Due to the constraints of hardware, memory, and power resources, certain IoT sensors cannot meet the corresponding computational requirements \cite{117}. The data exchange between IoT sensors and the network server also incurs communication overhead which scales up with the task sizes. Thus, new hardware design is needed toward future smart and powerful IoT devices. For instance, the work in \cite{118} introduces a software-based DL accelerator to support data training on mobile sensor hardware. \textcolor{black}{The key idea is to use a set of heterogeneous processors (e.g., GPUs) where each computing unit exploits distinct computational resources for processing different inference phases of DL models.} This aims to optimize hardware usage for data training without compromising the accuracy performances, enabled by two algorithms, i.e., runtime layer compression and deep architecture decomposition. Simulation results demonstrate the superior performance of the proposed approach in terms of low execution time and energy consumption in AI hardware running, compared to cloud offloading-based approaches. This research is promising to develop mobile AI inference for edge intelligence, which would enable on-device IoT implementation in the future wireless networks. \textcolor{black}{Moreover, a scheme called Tiny-transfer learning (TinyTL) is considered in \cite{302} for memory-efficient on-device sensor learning. To compensate for the capacity loss, a memory-efficient bias module, called lite residual module is integrated which enhances the model capacity by refining the intermediate feature maps of the feature extractor with a minimal memory overhead. Numerical simulations are implemented using image classification datasets, showing that the proposed on-device learning approach can achieve a competitive accuracy performance (above 90\%), compared to the traditional training solutions (e.g., Inception-V3), while reducing the training memory footprint by up to 12.9. \textcolor{black}{More research efforts are needed to provide hardware-based AI training solutions on nano IoT devices and embedded wearables in future intelligent 6G IoT networks, such as intelligent-enhanced living assistance services.}}
\subsection{Standard Specifications for 6G-IoT}
The emergence of 6G technologies potentially transforms the shape of IoT markets and revolutionizes the IoT ecosystems with advanced wireless networking features. However, the development of 6G-IoT systems requires stringent standard specifications that calls the collaboration of all business stakeholders such as network operators, services providers, and customers \cite{119}.  The lack of system standards can hinder the deployment of 6G functions and technologies in customer IoT systems.
Moreover, the introduction of vertical 6G-IoT use cases in future intelligent networks imposes major architectural changes to current mobile networks in order to simultaneously support a diverse variety of stringent requirements (e.g. autonomous driving, e-healthcare, etc.) In such a context, network standards hold an important role in deploying 6G-IoT ecosystems at large-scale due to the reliance of other important services such as computing and 6G server-IoT device communication protocols. \textcolor{black}{\textcolor{black}{A popular protocol is MODBUS \cite{303} which is the communication standard for connectivity of computer servers, industrial electronic equipments, and sensor devices in IoT environments. MODBUS is developed based on a variety of enabling protocols such as remote terminal unit (RTU), TCP/IP, and UDP. It relies on mesh networking architectures and can provide industrial communications with the supervisory control over the industrial radio bands.} Recently, the Industry Specification Group  of the European Telecommunications Standards Institute  has released the initiative called ETSI Multi-access Edge Computing \cite{120}, which aims to leverage seamlessly edge computing and communication frameworks for integrating various edge-based IoT applications originating from the vendors and service providers in the next-generation wireless networks. This would facilitate various IoT services such as video analytics, augmented reality, data caching, and content delivery.} In the near future, the interested stakeholders should pay more attention to develop new standard specifications for new space-air-ground-underwater communications, e.g., new IoT satellite IoT communications, which would be significant to the future deployment of new commercial IoT applications such as space travel and deep sea marine services. The summary of challenges and potential directions in 6G-IoT research is presented in Table~\ref{Table:Challenges}.

\begin{table*}
	\centering
	\caption{Summary of key research challenges and possible directions for 6G-IoT.}
	\label{Table:Challenges}
	\setlength{\tabcolsep}{5pt}
	\begin{tabular}{|P{3cm}|P{7cm}|P{7cm}|}
		\hline
		{\cellcolor{blue!25}\centering \textbf{Challenges}}& 	
		{\cellcolor{blue!25}\centering \textbf{Description}}& 
		{\cellcolor{blue!25}\textbf{Possible directions}	}
		\\
		\hline
		\multirow{10}{*}{\makecell{Security and Privacy\\ in 6G-IoT}}  &	\begin{itemize}
			\item	The developments of 6G-IoT come at the cost of new security and privacy concerns, e.g., unauthorized access to data at computing units/servers, threats to integrity in the access network infrastructure, and data breaches in  edge intelligence \cite{109}.
			\item	The deployment of satellite-UAV-IoT communications over the untrusted environments in the space can be hindered by data privacy leakage caused by third-parties and adversaries during the data exchange and transmission.
		\end{itemize} &	
		\begin{itemize}
			\item Perturbation techniques such as differential privacy or dummy can be useful to protect training datasets against data breach in the edge intelligence-based 6G-IoT networks \cite{111}.
			\item	Blockchain is a promising solution to build trust and establish secure decentralized communications for 6G-IoT networks \cite{113} over the space and untrusted wireless environments. 
		\end{itemize}
		\\
		\hline
		\multirow{8}{*}{\makecell{Energy Efficiency \\ in 6G-IoT}} &	\begin{itemize}
			\item	In future 6G-based IoT networks, how to achieve high energy efficiency is a major concern, e.g., energy resources needed for data transmission, communications, and service delivery services. 
			\item	Building energy-efficient wireless communication protocols is highly needed for green 6G-IoT networks.
			
		\end{itemize} &	
		\begin{itemize}
			\item The \cite{115} jointly optimizes the QoS and energy consumption in 6G-based smart automation systems, by implementing a 6G-driven multimedia data structure approach. 
			\item	Energy harvesting techniques to exploit the renewable energy resources will be very useful to build green 6G-IoT systems \cite{116}.  
		\end{itemize}
		\\
		\hline
		\multirow{8}{*}{\makecell{Hardware Constraints \\ of IoT Devices}} &	\begin{itemize}
			\item	The participations in 6G communications and computation tasks poses new challenges into hardware designs for IoT devices.
			\item	Due to the constraints of hardware, memory, and power resources, certain IoT sensors cannot meet these computational requirements in 6G-based customer applications \cite{117}. 
		\end{itemize} &	
		\begin{itemize}
			\item The work in \cite{118} introduces a software-based deep learning accelerator to support AI/DL training on mobile sensor hardware.
			\item In the future, it is desired to develop lightweight on-device hardware platforms to meet service computation demands, e.g., on-device edge intelligence in mobile 6G networks. 
		\end{itemize}
		\\
		\hline
		\multirow{9}{*}{\makecell{Standard Specifications \\ in 6G-IoT}} &	\begin{itemize}
			\item	The development of 6G-IoT systems requires stringent standard specifications that calls the collaboration of all business stakeholders such as network operators, services providers, and customers \cite{119}.
			\item		The introduction of vertical 6G-IoT use cases in future intelligent networks imposes major architectural changes to current mobile networks for supporting a diverse variety of stringent requirements.
		\end{itemize} &	
		\begin{itemize}
			\item It is important to establish new standards for computing and 6G server-IoT device communication protocols.
		\item  A standard initiative ETSI Multi-access Edge Computing is introduced in \cite{120}, allowing for leveraging seamlessly edge computing and communication resources in various edge-based IoT applications. 
		\end{itemize}
		\\
		\hline
	\end{tabular}
\end{table*}

\section{Conclusions}
\label{concl}
6G has recently sparked much interest in both industry and  academia due to its appealing features compared to previous generations of wireless networks. In this article, we have explored the opportunities brought by the 6G technologies to support IoT networks through a holistic survey based on the emerging study activities  in the field. This work is motivated by the lack of a comprehensive survey on the use of 6G for IoT. \textcolor{black}{To bridge this gap, we have first introduced the recent advances in FL and IoT and discussed the key requirements of 6G-IoT integration.} We have then identified and analyzed the key 6G technologies for enabling IoT networks, ranging from edge intelligence, RISs, space-air-ground-underwater communications, THz communications to mURLLC and blockchain. Next, we have provided a holistic discussion on the use of 6G in emerging IoT applications, such as HIoT, VIoT and autonomous driving, UAVs, SIoT, and IIoT. From the extensive survey, the key technical aspects and emerging use cases in 6G-IoT have been also summarized and analyzed via taxonomy tables. Finally, we have identified potential  challenges and highlighted possible directions for future research.  

Research on 6G-IoT networks and applications is still in its infancy. This being said, it is envisioned that 6G will transform the current IoT network infrastructures and bring new levels of service quality and user experience in the future applications. We believe our timely work will shed valuable light on the research of the 6G-IoT integration topics as well as motivate researchers and stakeholders to augment the research efforts in this promising area.

\balance
\bibliography{Ref}
\bibliographystyle{IEEEtran}

\end{document}